\documentclass{emulateapj}
\usepackage{apjfonts}
\usepackage{epsfig,color}
\usepackage{mathrsfs}
\usepackage{rotating}

\journalinfo{The Astrophysical Journal, 2012 in press}
\slugcomment{Received 2011 September 29; accepted 2011 December 16}
\shorttitle{Star formation and Formation of Broad Line Regions}
\shortauthors{WANG ET AL.}

\def\civ{C {\sc iv}}
\def\cala{{\cal A}}

\def\calc{{\mathscr C}}

\def\cald{{\cal D}}
\def\calh{{\mathscr H}}
\def\call{{\mathscr L}}

\def\calS{{\cal S}}

\def\dotmbh{\dot{M}_{\bullet}}

\def\elledd{\ell_{\rm Edd}}

\def\feii{{Fe {\sc ii}}}
\def\gbh{g_{\bullet}}
\def\hatn{{\hat{n}}
\def\hbeta{H\beta}}
\def\hcas{H_{_{\rm CAS}}}
\def\kappas{\kappa_{_{\rm S}}}

\def\mcas{M_{_{\rm CAS}}}

\def\mgii{Mg {\sc ii}}

\def\nh{N_{\rm H}}

\def\ncl{n_{\rm cl}}
\def\ncas{n_{_{\rm CAS}}}
\def\omegad{\omega_{\rm d}}

\def\ovii{O {\sc vii}}
\def\oviii{O {\sc viii}}

\def\tcas{T_{_{\rm CAS}}}
\def\tcl{T_{\rm cl}}

\def\rmK{{\rm K}}
\def\rout{R_{\rm out}}

\def\ergs{\ifmmode {\rm ergs~ s^{-1}} \else {\rm ergs~s^{-1}}\ \fi}
\def\kms{\ifmmode {\rm km~ s^{-1}} \else {\rm km~s^{-1}}\ \fi}

\def\mbh{M_{\bullet}}

\def\mgii{\ifmmode Mg {\sc ii} \else Mg {\sc ii}\ \fi}
\def\feii{\ifmmode Fe {\sc ii} \else Fe {\sc ii}\ \fi}
\def\ciii{\ifmmode C {\sc iii}] \else C {\sc iii}]\ \fi}
\def\heii{\ifmmode He {\sc ii} \else He {\sc ii}\ \fi}
\def\niii{\ifmmode N {\sc iii}] \else N {\sc iii}]\ \fi}
\def\nv{\ifmmode N {\sc v} \else N {\sc v}\ \fi}
\def\oiii{[O {\sc iii}]}
\def\oii{[O~{\sc ii}]}
\def\pp{\prime\prime}
\def\pcas{P_{\rm CAS}}

\def\sunm{M_{\odot}}
\def\sunmyr{M_{\odot}{\rm yr^{-1}}}

\def\apr{{\cal A}_{pR}}
\def\apz{{\cal A}_{pz}}
\def\arhor{{\cal A}_{\rho R}}
\def\arhoz{{\cal A}_{\rho z}}
\def\homega{\hat{\omega}}
\def\kr{k_R}
\def\kz{k_z}
\def\pp{\partial}

\def\pr{\partial R}
\def\prho{\partial \rho}
\def\pt{\partial t}
\def\pz{\partial z}

\def\vr{v_{_R}}
\def\vphi{v_{\phi}}
\def\vz{v_z}

\def\lax{{$\mathrel{\hbox{\rlap{\hbox{\lower4pt\hbox{$\sim$}}}\hbox{$<$}}}$}}
\def\gax{{$\mathrel{\hbox{\rlap{\hbox{\lower4pt\hbox{$\sim$}}}\hbox{$>$}}}$}}

\def\nvciv{N~{\sc v}/C~{\sc iv}}
\def\niiiciii{N~{\sc iii}]/C~{\sc iii}]}

\def\ergs{${\rm erg~s^{-1}}$}

\begin{document}

\title{Star formation in self-gravitating disks in active galactic nuclei.\\
II. Episodic formation of broad line regions}

\author{
Jian-Min Wang\altaffilmark{1,2},
Pu Du\altaffilmark{1},
Jack A. Baldwin\altaffilmark{3},
Jun-Qiang Ge\altaffilmark{1},
Chen Hu\altaffilmark{1},
and Gary J. Ferland\altaffilmark{4}
}

\altaffiltext{1}
{Key Laboratory for Particle Astrophysics, Institute of High Energy Physics,
Chinese Academy of Sciences, 19B Yuquan Road, Beijing 100049, China; email: wangjm@mail.ihep.ac.cn}

\altaffiltext{2}
{National Astronomical Observatories of China, Chinese Academy of Sciences, 20A Datun Road,
Beijing 100020, China}

\altaffiltext{3}
{Physics and Astronomy Department, 3270 Biomedical Physical Sciences Building, Michigan State University,
East Lansing, MI 48824}

\altaffiltext{4}
{Department of Physics and Astronomy, 177 Chemistry/Physics Building,
University of Kentucky, Lexington, KY 40506}


\begin{abstract}
This is the second in a series of papers discussing the process and effects of star formation in the
self-gravitating disk around the supermassive black holes (SMBHs) in active galactic nuclei (AGNs). We
have previously suggested that warm skins are formed above the star forming (SF) disk through the diffusion
of warm gas driven by supernova explosions. Here we study the evolution of the warm skins when they are
exposed to the powerful radiation from the inner part of the accretion disk. The skins initially are heated
to the Compton temperature, forming a Compton atmosphere (CAS) whose subsequent evolution is divided into
four phases. Phase I is the duration of pure accumulation supplied by the SF disk.
During phase II clouds begin to form due to line cooling and sink to the SF disk. Phase
III is a period of preventing clouds from sinking to the SF disk through dynamic interaction between
clouds and the CAS because of the CAS over-density driven by continuous injection of warm gas from the
SF disk. Finally, phase IV is an inevitable collapse of the entire CAS through line cooling.
This CAS evolution drives the episodic appearance of BLRs.

We follow the formation of cold clouds through the thermal instability of the CAS
during phases II and III, using linear analysis. Since the clouds are produced inside the CAS,
the initial spatial distribution of newly formed clouds and angular momentum naturally follow the
CAS dynamics, producing a flattened disk of clouds. The number of clouds in phases II and III can be
estimated, as well as the filling factor of clouds in the BLR. Since the cooling function depends
on the metallicity, the metallicity gradients that originate in the SF disk give rise to different
properties of clouds in different radial regions. We find from the instability analysis that clouds
have column density $\nh \lesssim 10^{22}{\rm cm^{-2}}$ in the metal-rich regions whereas they
have $\nh \gtrsim 10^{22}{\rm cm^{-2}}$ in the metal-poor regions. The metal-rich clouds compose the
high ionization line (HIL) regions whereas the metal-poor clouds are in low ionization line
(LIL) regions. Since metal-rich clouds are optically thin, they will be blown away by radiation
pressure, forming the observed outflows. The outflowing clouds could set up a metallicity correlation
between the broad line regions and narrow line regions. The LIL regions are episodic due to the mass
cycle of clouds with the CAS in response to continuous injection by the SF disk, giving rise to different
types of AGNs. Based on SDSS quasar spectra, we identify a spectral sequence in light of emission line
equivalent width from Phase I to IV. A key phase in the episodic appearance of the BLRs is bright type
II AGNs with no or only weak BLRs, contrary to the popular picture in which the absence of a BLR is due
to a low accretion rate. We discuss observational implications and tests of the theoretical predictions
of this model.
\end{abstract}
\keywords{black hole physics --- galaxies: evolution --- quasars: general}

\section{Introduction}
Accretion onto supermassive black holes liberates copious amounts of energy, mainly as big blue bumps
and X-rays, driving many phenomena associated with active galactic nuclei (AGNs), such as broad emission
lines and outflows. Numerous attempts have been made over the past four decades to construct a
self-consistent model of AGNs, but these have not yet been successful and the structure of AGNs and
quasars remains largely unsolved. Principle Component Analysis (PCA) for large samples of AGNs
(Boroson \& Green 1992; vanden Berk et al. 2001; Marziani et al. 2003; 2010) has shown that the
Eddington ratio is tightly correlated with eigenvector 1 and the properties of the broad emission
lines. In addition, the broad line region (BLR) metallicity relates with AGN luminosity or Eddington
ratios, linking with the feeding process (Hamann \& Ferland 1999; Shemmer et al. 2004; Warner et al. 2004;
Matsuoka et al. 2011). It is likely that accretion onto the super-massive black holes (SMBHs) and the
appearance of BLRs are accompanying processes resulting from supplying gas to the central engine, and the
properties of the BLR are determined by the supplying processes.

Spectra of Seyfert galaxies and quasars generally show broad emission lines with
full-width at half-maximum (FWHM) of a few $10^3\kms$ (see the reviews by Osterbrock \& Mathews 1986;
Netzer 1990; Sulentic et al. 2000 and Ho 2008). BLRs are generally discussed as being composed
of discrete rapidly moving clouds responsible for emitting the observed broad lines, although the
smoothness of the observed profiles indicates that there must be large numbers of such clouds (Arav et al. 1997;
 Ferland 2004; Laor 2006; Laor et al. 2006).

\begin{figure*}[t!]
\centering
\figurenum{1}
\includegraphics[angle=-90.0,scale=1.0]{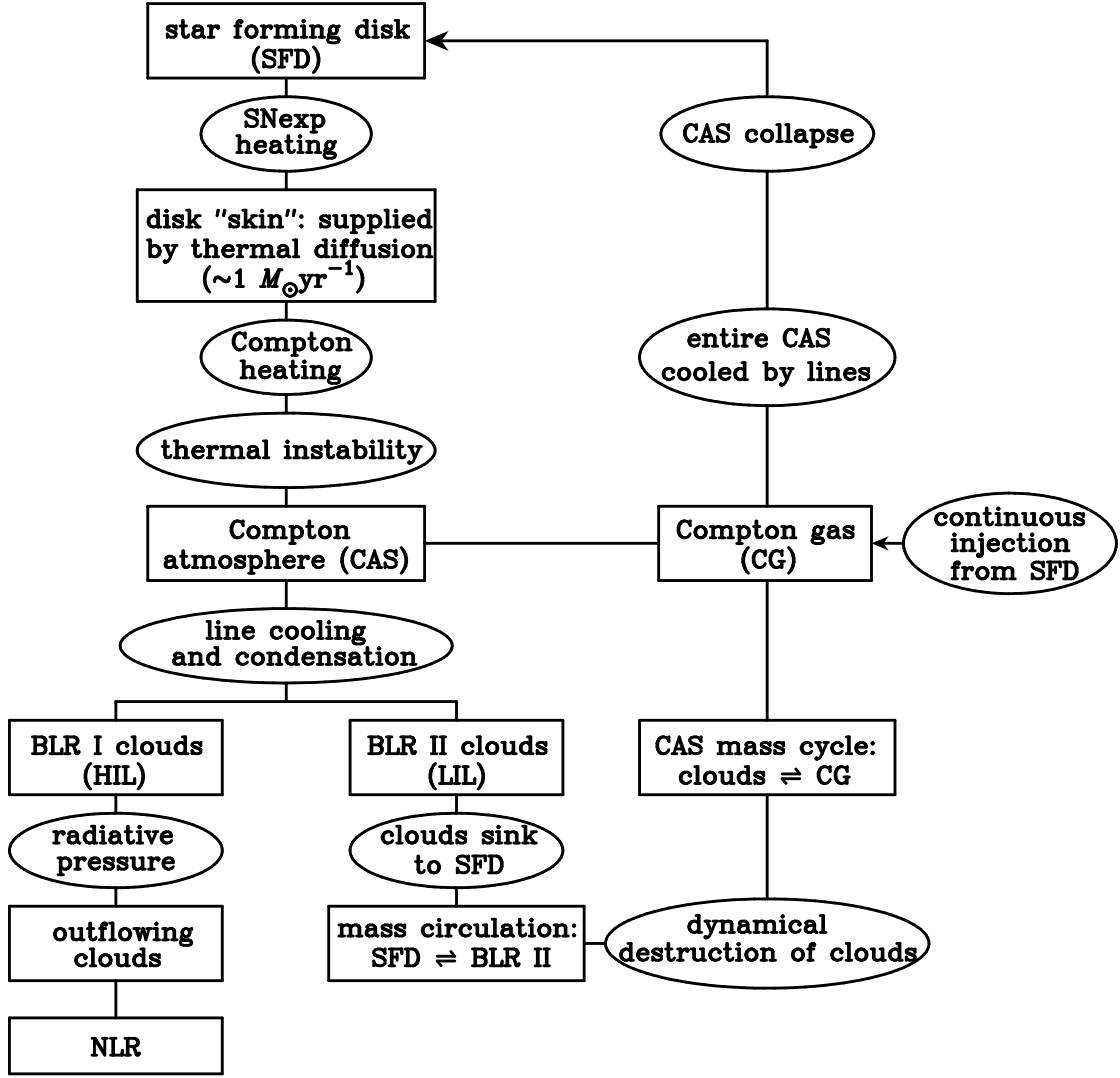}
\figcaption{\footnotesize Flow chart of the model of episodic formation of broad line regions presented in this paper.
It starts with the SF disk, where evaporation is driving thermal diffusion of hot gas through SNexp.
The diffuse gas then undergoes various physical processes, yielding broad line regions as one
of the by-products of feeding the SMBHs. The ellipses represent the underlying physics whereas the
rectangles show states of the gas. The symbol $``\rightleftharpoons"$ indicates a cycle or a circulation of mass. See
details in the main text of the paper.}
\label{chart_flow}
\end{figure*}

It has been speculated for a long time that these clouds are formed from the thermal instability of
the hot gas. However, some major questions remain open:
\begin{itemize}
\item{}Why does the BLR form?
\item What are the origins of the emitting gas and its high metalicity?
\item What are properties of cold clouds formed by thermal instability?
\item What is the geometry of the BLR?
\item Do the BLR clouds relate to phenomenon of AGN outflows?
\end{itemize}

The existing models listed below in \S6 generally treat the above issues as being independent of each
other, but there is increasing evidence that they are in fact closely related. First, the
gravitational energy is mainly released around $3-100$ Schwarzschild radii, but the well-known
reverberation mapping relation of the strong correlation between BLR size and ionizing luminosity
indicates $R_{\rm BLR}\propto L^{0.5}$ with $ R_{\rm BLR} \sim 10^4$ Schwarzschild radii (Kaspi et al.
2000; 2005). The mapping relation is so strong that we believe there is an intrinsic connection between
the accretion onto the SMBHs and BLR formation though the two scales are differing by a factor of
$\sim 10^2$. The intrinsic connection could be complex somehow. If the broadening mechanism is due
to an assembly of clouds with Keplerian rotation, the differences in width of different
emission lines indicates that the BLR is an extended region. A flattened disk is then favored
(Laor 2007; Collin et al. 2006). Second, it is well-known that quasars are metal-rich,
generally a few times the solar abundance and up to ten or more times the solar value (Hamann \&
Ferland 1992), and that higher metallicity correlates with higher luminosity, Eddington ratios or SMBH
mass (Hamann \& Ferland 1999; Shemmer et al. 2004; Netzer et al. 2004; Warner et al. 2004; Matsuoka et
al. 2011). In particular, the metallicity does {\em not} relate with the IR luminosity (Simon \& Hamann
2010). All of these results provide  potential clues for understanding the formation of BLRs and related
phenomena, in the context of the BLR being a by-product of the feeding process. In addition, it has
gradually been realized that star formation is not only important for feeding
the SMBHs, but also for the metallicity observed in the broad line regions (Wang et al. 2010; 2011).

\begin{deluxetable*}{ll}
\tabletypesize{\footnotesize}
\tablewidth{0pt}
\tablecaption{{\sc Summary of Terminology Used in the Model}}
\tablehead{
Phrase & Physical meanings}
\startdata
SMBH accretion disk     & accretion flows within the self-gravitating radius ($R_{\rm SG}$)\\
star forming disk (SFD) & accretion flows between $R_{\rm SG}$ and the inner edge of dusty torus\\	
warm gas                & gas heated to $\sim 10^6$K by the SNexp inside the SF disk \\
warm skin               & warm gas bound by the SMBH potential\\
Compton atmosphere (CAS)& is composed of Compton gas and cold clouds\\
Compton gas (CG)        & hot gas at $T_{\rm Comp}$ heated by AGN and with $\Xi_1\le\Xi\le \Xi_2$ (see \S2.2.4)\\
cold clouds             & clouds in the CAS that are formed through thermal instability\\
outflowing clouds       & clouds driven by radiation pressure to move radially outward
\enddata
\label{summary_words}
\end{deluxetable*}

The main goal of this paper is to make detailed predictions using a self-consistent model which addresses
all of these aspects. Figure \ref{chart_flow} sketches the global scenario of the model of BLR formation
and evolution that we will develop and use here. For convenience, Table \ref{summary_words} lists the
terminology used in this paper. Evaporation of molecular clouds driven by supernova explosions (SNexp) in
the star-forming (SF) disk plays a key role in BLR formation. The gas evaporated by SNexp has a temperature
of $10^6$K, but is still bound by the SMBH potential, forming a warm ``skin" of the SF disk. Emission from
the accretion disk around the SMBH heats the warm skin and establishes a ``Compton atmosphere" (CAS) above
the SF disk. The CAS is composed of cold clouds and Compton gas, whose fate is then determined by the interplay
between Compton heating, cooling and thermal instability, and by angular momentum redistribution, determining
the BLR geometry and dynamics. The interaction between the clouds and the Compton gas, along with the continuous
supply of warm gas from the SF disk, causes the clouds to have complicated lives, giving rise to an episodic
appearance of the BLRs. The advantages of the present model result from a natural assembly of a series of
physical processes, providing a self-consistent solution that includes feeding the SMBH, metal production,
and BLR formation.

In a previous paper (Wang et al. 2011a, hereafter W11), we showed that SNexp in the SF disk will expel
gas that will then form a warm skin above the disk. This current paper continues on from that point,
and is structured as follows. \S2 gives basic considerations for the underlying physical processes.
\S3 is devoted to discussing the fate of the atmosphere of the warm gas evaporated from the SF disk,
and we find that there are four phases of the evolution driving the appearance of different types
of AGNs. \S4 discusses the high and low ionization line regions. Observational tests of the present
model are extensively discussed in \S5. In \S6, we present a brief comparison of the current model
with other existing models, and point out future work that is needed. Finally, in \S7,  we summarize
our conclusions.

\section{Basic considerations}
This section is devoted to basic considerations of the underlying physical processes of the
atmosphere above the SF disk. After warm skins have inevitably developed above the SF disk
as a result of the blast waves of SNexp (W11), they are exposed to the intensive radiation
field of the accretion disk and are heated to produce ascending Compton gas. It should
be emphasized that the skins are continuously supplied by the SF disk, resulting in a
non-stationary state in the Compton gas. In addition, the rotating Compton gas will undergo
thermal instability once its density is high enough, as well as having a dynamical interaction
with the SMBHs. These competitive processes determine the basic properties of the BLR.

The SF disk is much larger than the $3-100R_{\rm Sch}$ region where most gravitational energy
is released, where $R_{\rm Sch}=2G\mbh/c^2=3.0\times 10^{13}~M_8$ cm is the Schwarzschild radius, $G$
is the gravity constant, $\mbh=10^8M_8\sunm$ is the SMBH mass, and $c$ is the light speed. For
simplicity, we assume that the radiation field of the accretion disk is isotropic as a point energy
source and that the star forming disk has zero thickness.

%

\begin{figure*}
\centering
\figurenum{2}
\includegraphics[angle=-90,scale=0.58]{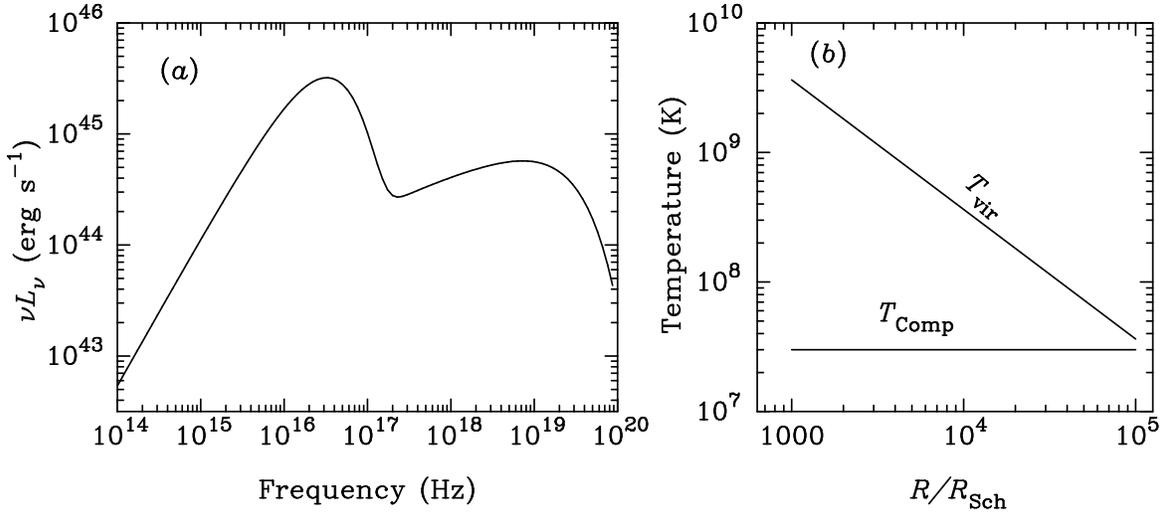}
\figcaption{\footnotesize
({\em a}): SED from accretion disk with corona. We use the emergent spectrum $F_\nu\propto \nu^{1/3}$ for
optical-soft X-rays from the Shakura-Sunyaev disk. The spectrum from soft to hard X-rays is a power-law with a
cutoff. The bolometric luminosity is $10^{46}$\ergs. ({\em b}):
The Compton temperature ($T_{\rm Comp}\sim 3\times 10^3$K from the SED given by the panel {\em a}) in
the surroundings around the SMBHs, compared with the virial temperature. Over most of the
region within 1pc, $ T_{\rm Comp}$ is lower than $ T_{\rm vir}$, implying that all the Compton gas is bound by the SMBH potential.}
\label{sed_T}
\end{figure*}

\subsection{Gas supply from the SF disk: warm skins}
It is convenient to review some results from W11. Typical values of the parameters describing the
SF disk are given in table 1 of W11. The blast
waves of SNexp interact with molecular clouds and evaporate warm gas (McKee \& Cowie 1975).
The SF disk is assumed to be located in the region between the self-gravitating radius ($R_{\rm SG}$)
and the inner edge of the torus. The self-gravitating radius is given by
$R_{\rm SG}\approx 511~\alpha^{2/9}M_8^{-2/9}\dot{m}^{4/9} R_{\rm Sch}$ (Laor \& Netzer 1989),
where $\alpha$ is the viscosity parameter, $\dot{m}=\dot{M}/\dot{M}_{\rm Edd}$,
$\dot{M}_{\rm Edd}=2.22\eta_{0.1}^{-1}M_8~\sunmyr$,
and $\eta_{0.1}=\eta/0.1$ is the radiative efficiency, assuming Toomre's parameter
$Q=1$ (Goodman 2003; Rafikov 2009; Collin \& Zahn 2008). We take as the inner radius of the dusty
torus $R_{\rm torus}=1.0$pc, for quasars with $\mbh=10^8\sunm$.
The density and temperature of the warm gas inside the SF disk are determined by the
balance of heating and cooling. W11 give the density and temperature of the hot plasma in the SF disk as
\begin{equation}
n_6=0.88~f_{\xi,-3}^{0.65}E_{\rm 51}^{0.65}\mu_0^{0.45}Z_0^{-0.36}\dot{\Sigma}_{*,0}^{0.65}H_{14}^{-0.36},
\end{equation}
and
\begin{equation}
T_6=1.40~f_{\xi,-3}^{0.23}\mu_0^{0.69}E_{51}^{0.23}Z_0^{0.23}\dot{\Sigma}_{*,0}^{0.23}H_{14}^{0.23}.
\end{equation}
where $n_6=n_e/10^6{\rm cm^{-3}}$, $T_6=T_e/10^6$K, $Z_0=Z/Z_{\odot}$ is the metallicity,
$E_{51}=E_{\rm SN}/10^{51}{\rm erg}$ is the SNexp energy, $\dot{\Sigma}_{*,0}$ is the surface density of
star formation rate in units of $\sunm {\rm yr^{-1} pc^{-2}}$, $H_{14}=H_{\rm d}/10^{14}{\rm cm}$ is the
thickness of the SF disk and $f_{\xi}$ is a parameter related with the initial mass function in units of $10^{-3}$.
The sound speed of the warm gas is $c_s=10^7T_6^{1/2}{\rm cm~s^{-1}}$, which is smaller than the
escape velocity $V_{\rm esc}=4.2\times 10^8~r_4^{-1/2}{\rm cm~s^{-1}}$ (within $10^4R_{\rm Sch}$).
Therefore, the warm gas is still bound by the SMBH potential, forming warm skins.
The mass rate of warm gas injection through diffusion is then given by
\begin{equation}
\begin{array}{lll}
\dot{M}_{\rm diff}\!&=\!&{\displaystyle \int_{R_{\rm in}}^{R_{\rm out}}2\pi c_sm_pn_eR_{\rm D} dR_{\rm D}}\\
\!&=\!&\displaystyle 1.8
\int_{x^{\prime}_{_{\rm in}}}^{x^{\prime}_{_{\rm out}}}f_{\xi,-3}^{0.77}E_{51}^{0.77}Z_0^{-0.24}\mu_0^{0.31}
\dot{\Sigma}_{*,0}^{0.77}H_{14}^{-0.24}x^{\prime} dx^{\prime}~\sunmyr,
\end{array}
\end{equation}
where $R_{\rm D}$ is the radius of the SF disk, $x^{\prime}=R_{\rm D}/10^{-0.5}{\rm pc}$,
$R_{\rm in}=R_{\rm SG}$ and $R_{\rm out}=1$pc. We find that the total mass rates of the warm gas diffusion
over the entire SF disk can be as high as a few $\sunmyr$, which exceeds the Eddington limit
for a $10^8\sunm$ SMBHs. It should be noted that the diffusion rates are much smaller than the inflow rate
of the SF disk ($\sim 10\sunmyr$; see Wang et al. 2010), and the skins have less significant influence on
the SF disk itself.

The warm skins caused by the evaporation from the SF disk have the same angular momentum as the disk. The
thickness of the warm skin is given by the vertical static equilibrium condition,
\begin{equation}
H_{\rm skin}=\frac{c_s}{\Omega_{\rm K}}\approx 1.29\times 10^{16}~T_6^{1/2}M_8r_4^{3/2}~{\rm cm},
\end{equation}
where $\Omega_{\rm K}=\left(G\mbh/R^3\right)^{1/2}$ is the Keplerian velocity,
$r_4=R/10^4R_{\rm Sch}$ and $T_6=T_e/10^6$K. The relative height of the skin is
$H_{\rm skin}/R\approx 0.05~T_6^{1/2}r_4^{1/2}$, indicating that the skin is flattened.
Since the skins are exposed to the
central engine, they are undergoing AGN heating and expansion after dynamical adjustment.

The thermal diffusion of warm gas from the SF disk will continue as long as the pressure
of warm gas in the disk is larger than the pressure of the Compton atmosphere. The warm gas pressure is given by
\begin{equation}
P_{\rm SF}=1.7\times 10^{-4}~f_{\xi,-3}^{0.88}\mu_0^{1.14}E_{51}^{0.88}Z_0^{-0.13}
           \dot{\Sigma}_{*,0}^{0.88}H_{14}^{-0.13}~{\rm dyn~cm^{-2}}.
\end{equation}
This equation indicates that $P_{\rm SF}$ is sensitive to the star formation rates, and insensitive
to the metallicity and height of the disk. This pressure determines the mass of the BLR gas and its
structure. We will show that the continuous injection driven by this pressure leads to a transient
appearance of the BLR. In our discussion of the BLR formation, we assume a steady star formation
rate in the SF disk during one episode of SMBH activity.

We would like to emphasize that the properties of the warm skin are governed by the star formation rate in the SF disk,
or by the surface density of the disk itself. Since the skins are totally influenced by the central engine,
they just provide the boundary condition of the Compton atmosphere for the BLR formation
discussed later. The CAS undergoes a complicated evolution which drives the episodic appearance
of a BLR even in response to continuous injection from the SF disk.

\begin{figure*}
\centering
\figurenum{3}
\includegraphics[angle=-90,scale=0.65]{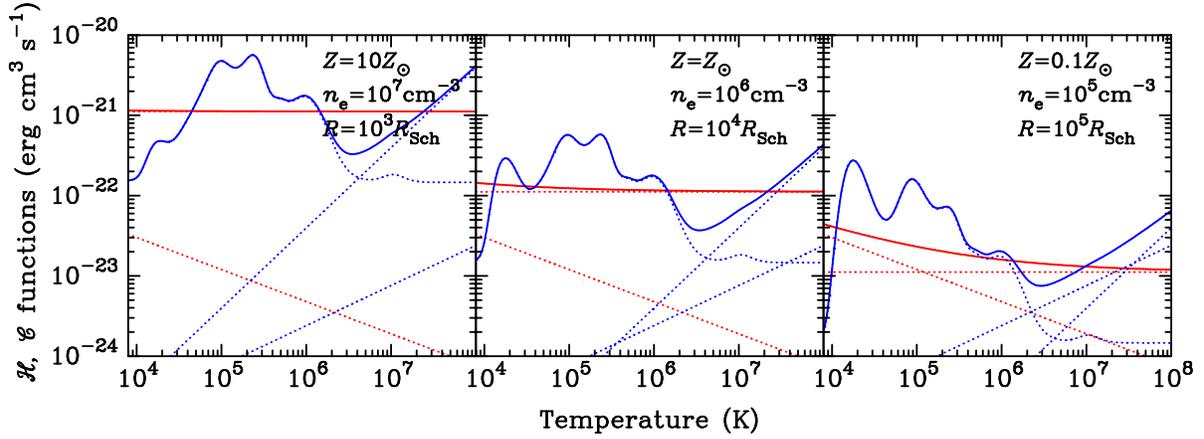}
\figcaption{\footnotesize
The heating and cooling functions versus temperature and metallicity at different radii. The blue dotted
lines are cooling functions ($\calc_{\rm Comp}$, $\calc_{\rm ff}$ and $\calc_{\rm line}$) and the red
dotted lines are heating functions ($\calh_{\rm Comp}$ and $\calh_{\rm ph}$). The red and blue solid lines
are the total heating and cooling functions, respectively. We set the metallicity $Z=10, 1, 0.1Z_\odot$ at
different radii, and typical densities are consistent with results shown by Figure 6,
respectively.
}
\label{HC}
\end{figure*}

\subsection{Heating and cooling functions}
\subsubsection{Heating}
We mainly consider two processes of heating by the AGN continuum as the main mechanisms: Compton heating and
photoionzation. Assuming a point energy source with luminosity $L_{\nu}$ ($=4\pi R^2F_{\nu}$) at the
center, Compton heating rates are given by (Levich and Sunyaev 1970)
\begin{equation}
\begin{array}{ll}
\calh_{\rm Comp}&=\displaystyle \frac{2\pi\sigma_{\rm T}\hbar}{m_ec^2}\frac{1}{n_e}\int \nu F_{\nu} d\nu, \\
                &=4.90\times 10^{-28}~n_6^{-1}r_4^{-2}M_8^{-2}\\
                &\qquad\quad\times\displaystyle\int_{10^{14}{\rm Hz}}^{\rm 100keV}
                 \left(\nu L_{\nu}\right)_{45}d\nu_{14}~~{\rm erg~cm^3~s^{-1}},
                 \end{array}
\end{equation}
where $F_{\nu}$ is the photon flux of the AGN continuum (in ${\rm erg~s^{-1}~Hz^{-1}~cm^{-2}}$),
$\sigma_{\rm T}=6.65\times 10^{-25}{\rm cm^2}$ is the Thompson cross section,
$\hbar=1.06\times 10^{-27}{\rm erg~s}$ is the Planck constant, $m_e$ is the mass of the electron
and $n_e$ is the number density of electrons. Here
$(\nu L_\nu)_{45}=\nu L_\nu/10^{45}$\ergs, $r_4=R/10^4R_{\rm Sch}$, $R$ is the distance of the
ionized gas to the center and $\nu_{14}=\nu/10^{14}$Hz. Here the Compton heating is simply expressed
for a CAS with a constant density typical of that at the characteristic radius. We include the
dependence on radius in following calculations. The heating rate due to photoionization is given by
\begin{equation}
\calh_{\rm ph}\approx 3.01\times 10^{-23}~T_4^{-0.4}~~{\rm erg~cm^3~s^{-1}},
\end{equation}
where $T_4=T/10^4$K (Beltrametti 1981). We note that the photoionization heating rate must actually
depend on the ionization parameter (defined by equation 13). Here we just use a simple approximation
from the published literature that is roughly correct for the conditions in the BLR. Fortunately,
Equation (7) does not affect the triggering thermal instability, although the final temperature of
cold clouds depends on the ionization parameter. This simplified version of the heating will be improved
in a future paper. The free-free absorption coefficient is
$\tilde{\kappa}_{\rm ff}=1.32\times 10^{-27}T_4^{-3/2}\nu_{10}^{-2}~~{\rm cm^{-4}}$, where
$\nu_{10}=\nu/10^{10}{\rm Hz}$ (Rybicki \& Lightman 1979), and only becomes significant at radio
frequencies. Considering that the gravitational energy is originally released from optical to hard
X-rays, the free-free absorption is neglected here though radio and IR photons can significantly heat
the medium through this process (Ferland \& Baldwin 1999). However,
for those low luminosity AGNs powered by advection-dominated accretion flows, heating by radio emission
cannot be neglected in the total energy budget of emission lines. In particular, clouds responsible
for emission lines in Low Ionization Emission Regions (LINERs) have been strongly influenced by the
radio heating. However, we focus here on the radio-quiet quasars in which the radio emission is much
fainter than the other bands.  We have the total heating function as
$\calh=\calh_{\rm Comp}+\calh_{\rm ph}$.

\subsubsection{Cooling}
Three cooling processes are important: 1) Compton cooling; 2) free-free cooling and
3) line cooling. The Compton cooling is given by
\begin{equation}
\begin{array}{ll}
\calc_{\rm Comp}&=\displaystyle \frac{4kT\sigma_{\rm T}}{m_ec^2n_e}\int F_{\nu}d\nu,\\
                &=3.90\times 10^{-27}~~r_4^{-2}M_8^{-2}n_6^{-1}T_4\\
                &\qquad\qquad\quad\times\displaystyle
                \int_{10^{14}{\rm Hz}}^{100{\rm keV}}L_{\nu,31}d\nu_{14}
                ~~{\rm erg~cm^3~s^{-1}},
                \end{array}
\end{equation}
where $L_{\nu,31}=L_{\nu}/10^{31}{\rm erg~s^{-1}~Hz^{-1}}$, $\nu_{14}=\nu/10^{14}$Hz, and the
bremsstrahlung cooling is
\begin{equation}
\calc_{\rm ff}=2.4\times 10^{-27}T^{1/2}=2.4\times 10^{-25}~T_4^{1/2}~~{\rm erg~cm^3~s^{-1}}.
\end{equation}
The line cooling function has been extensively studied by many authors (e.g. B\"ohringer \& Hensler
1989; Sutherland \& Dopita 1993; Gnat \& Sternberg 2007). It is mainly characterized by three bumps,
which are contributed by the elements H, He, (C, N, Si, S, Ne, Mg, O) and H-like Fe ions, respectively.
We fit each of these peaks by parabolic curves as approximations of the line cooling functions
shown in Figure 2 in B\"ohringer \& Hensler (1989). The line cooling function for a plasma with the
abundance $Z$ is
\begin{equation}
\calc_{\rm line}=\sum_i 10^{f_i}+\left(\frac{Z}{Z_{\odot}}\right)\sum_{j}10^{f_j}
               ~~{\rm erg~cm^3~s^{-1}},
\end{equation}

where $i={\rm H,He}$, $j={\rm C,O,Ne,Fe, Fe^{25+}}$. We set a form of the cooling function as
$f_{k}=a_1+a_2\exp\left[-\left(\log T-b_1\right)^2/b_2\right]+a_3\exp\left[-\left(\log T-c_1\right)^2/c_2\right]$,
where $k=i,j$, to fit the numerical cooling functions given by B\"ohringer \& Hensler (1989). The coefficients
for different elements are listed by a sequence of ($a_1,a_2,a_3,b_1,b_2,c_1,c_2$). We have
H: (-59.37, 36.21, 6.47, 4.97, 4.93, 4.01, 0.26); He: (-30.34, 0.81, 7.65, 4.90, 0.056, 5.11, 3.03);
C: (-24.64, 2.59, 0.79, 4.77, 0.42, 5.03, 0.052); O: (-23.00, 1.39, 0.96, 5.11, 0.15, 5.44, 0.043)
Ne: (-23.58, 1.13, 1.29, 5.43, 0.051, 5.73, 0.044); Fe: (-25.21, 1.26, 2.59, 0.013, 81.49, 5.99, 0.35);
Fe$^{25+}$: (-23.79, 0.51, -0.032, 7.03, 0.047, 3.34, 15.81) for a plasma with solar abundances.
This approximation is accurate to within 10\% over the whole domain of temperatures from $10^4-10^7$K, but it
does not apply free-free emission of fully ionized gas ($>10^8$K). This is sufficient for the purposes of the
present paper. The total cooling functions is then $\calc=\calc_{\rm Comp}+\calc_{\rm ff}+\calc_{\rm line}$.

We wish to stress the important role of metallicity in the cooling. The metallicity gradient suggested
by W11 implies different cooling functions at different radii. As shown by subsequent Figure 3, the
inner regions
are overheated, if they are metal-poor, so that cold clouds are forbidden to form. A proper metallicity
is necessary at a radius to cool the heating so that formation of clouds is permitted at this distance.
The gradients give rise to different properties of cold clouds as a function of distance from the black
hole.

\subsubsection{Compton temperature}
The SEDs of radio-quiet quasars and AGNs are characterized by two humps\footnote{The present
$\langle \epsilon\rangle$ excludes the additional humps at infrared and radio wavelengths since they
do not originate in the central $1.0$pc region of interest here. Photons from the SF disk are also
neglected when determining the Compton temperature since they are dominated by photons from the accretion
disk.}: 1) the so-called big blue bump extending from optical to soft X-rays; and 2) hard X-rays with a
cutoff at about 100 keV (e.g. Richards et al. 2006; Vasudevan et al. 2009; Grupe et al. 2010). The Compton
temperature is determined by the photon mean energy ($\langle \epsilon\rangle$) of the SED in the $1.0$pc
region. The hard X-ray SED is given by $F_{\rm HX}(E)\propto E^{-\alpha_{\rm HX}}$, where
$\alpha_{\rm HX}=1.2\sim1.7$ with a cutoff at roughly $100 - 200$ keV (with the exception of the narrow
line Seyfert 1 galaxies in Grupe et al.'s (2010) sample, which have quite soft spectra in hard X-rays).
The optical-UV SED has index $\alpha_{\rm OUV}=1.5$ and a cutoff at 0.1keV. For this mean quasar SED,
the Compton temperature is
\begin{equation}
T_{\rm Comp}=\frac{1}{4k}\langle \epsilon\rangle={\rm a~few}\times 10^7~{\rm K},
\end{equation}
agreeing with Netzer (2008), where $k$ is the Boltzmann constant, in light of the SED. This Compton
temperature is at or slightly above the critical values below which serious absorption appears in soft
X-rays (e.g. Petre et al. 1984; Mathews
\& Ferland 1987). We use the Compton temperature $T_{\rm Comp}=3\times 10^7$K in this paper,
which is generally lower than the virial temperature given by
$T_{\rm vir}=3.63\times 10^{12}r^{-1}$K, indicating the diffuse gas is still bound by the potential
of the SMBHs. This heated gas forms an atmosphere above the SF disk.

The timescale of Compton heating is
\begin{equation}
t_{\rm Comp}=\frac{3m_ec^2}{4\sigma_{\rm T}}\frac{4\pi R^2}{\elledd L_{\rm Edd}}=\left\{
             \begin{array}{l}
             8.54\times 10^2~r_5^2\ell_{0.3}^{-1} M_8~{\rm yr},\\
                                               \\
              8.54~r_4^2\ell_{0.3}^{-1} M_8~{\rm yr},\\
                                               \\
              0.085~r_3^2\ell_{0.3}^{-1}M_8~{\rm yr},\end{array}\right.
\end{equation}
where $r_3=R/10^3R_{\rm Sch}$, $r_5=R/10^5R_{\rm Sch}$, $R$ is the distance from the SMBH,
$\ell_{0.3}=\elledd/0.3$ and $\elledd=L_{\rm Bol}/L_{\rm Edd}$ is the Eddington ratio and $L_{\rm Bol}$
is the bolometric luminosity. This timescale is very important
for determining the fate of the warm skins above the SF disk.

From Figure 3, we find that the Compton temperature of the CAS slightly increases with metallicity, that is to
say, it is a function of the radial distance from the black hole. From the figure, we find the approximate dependence
$T_{\rm Comp}\propto R^{-0.3}$. The CAS is not an exactly isothermal atmosphere. It should be pointed out that
formation of clouds is a local event in the CAS, and the global properties of the CAS hardly affect cloud formation.

\subsubsection{Ionization parameter}
The ionization parameter of the gas supplied by the SF disk is
\begin{equation}
\Xi=\frac{L_{\rm ion}}{4\pi R^2c n_ekT}=2.2~r_4^{-2}L_{45}(n_eT_e)_{14}^{-1}M_8^{-2},\\
\end{equation}
where $L_{45}=L_{\rm ion}/10^{45}$\ergs\,
is the ionizing luminosity and $(n_eT_e)_{14}$ is in units of $10^{14}{\rm cm^{-3}K}$. This definition
is very convenient for discussing the two-phase model with a pressure balance. It is well-known that the
S-shaped relation between $\Xi$ and $T$ shows the thermal states of the ionized plasma (Krolik et al. 1981).
When $\Xi\gtrsim 10$, the ionized plasma will have only one hot phase with Compton temperature.
Only a cold phase exists when $\Xi\lesssim 0.1$.

Figure \ref{sed_T}{\em b} shows a comparison of the Compton and virial temperatures,
indicating that the Compton gas is not able to escape from the SMBH potential. In the presence of
continuous injection due to thermal diffusion, the Compton gas accumulates and its density increases.
The ionization parameter of the Compton gas will drop until thermal instability develops. As we
argue below, a two-phase medium will form in the Compton gas once
the gas injected from the SF disk is sufficiently dense.

The ionization parameter is a strong function of the distance to the black hole. As a convenient estimate
of the thermal state of the ionized gas, we divide the regions into two parts with the boundary at $\Xi=1$,
corresponding to $R=10^4R_{\rm Sch}$. The inner part, known as the high ionization line (HIL) region, has
higher values of the ionization parameter with typical $\log \Xi=0.5$. The outer part, called the low
ionization line (LIL) region, is characterized here by the lower value $\log \Xi=-0.5$. Figure 4 illustrates
the HIL and LIL regions. Detailed discussions are given in Section 3.4.3, 4.2 and 4.3. Another definition
of ionization parameter as $U=\left(4\pi R^2 c n_e\right)^{-1}\int L_{\nu}d\nu/h\nu$ (Osterbrock \& Ferland
2006) is often used to distinguish the LIL and HIL regions, with $U>10^{-2}$ corresponding to the HIL
region(e.g. Marziani et al. 2010).

{
\centering
\figurenum{4}
\includegraphics[angle=-90,scale=0.35]{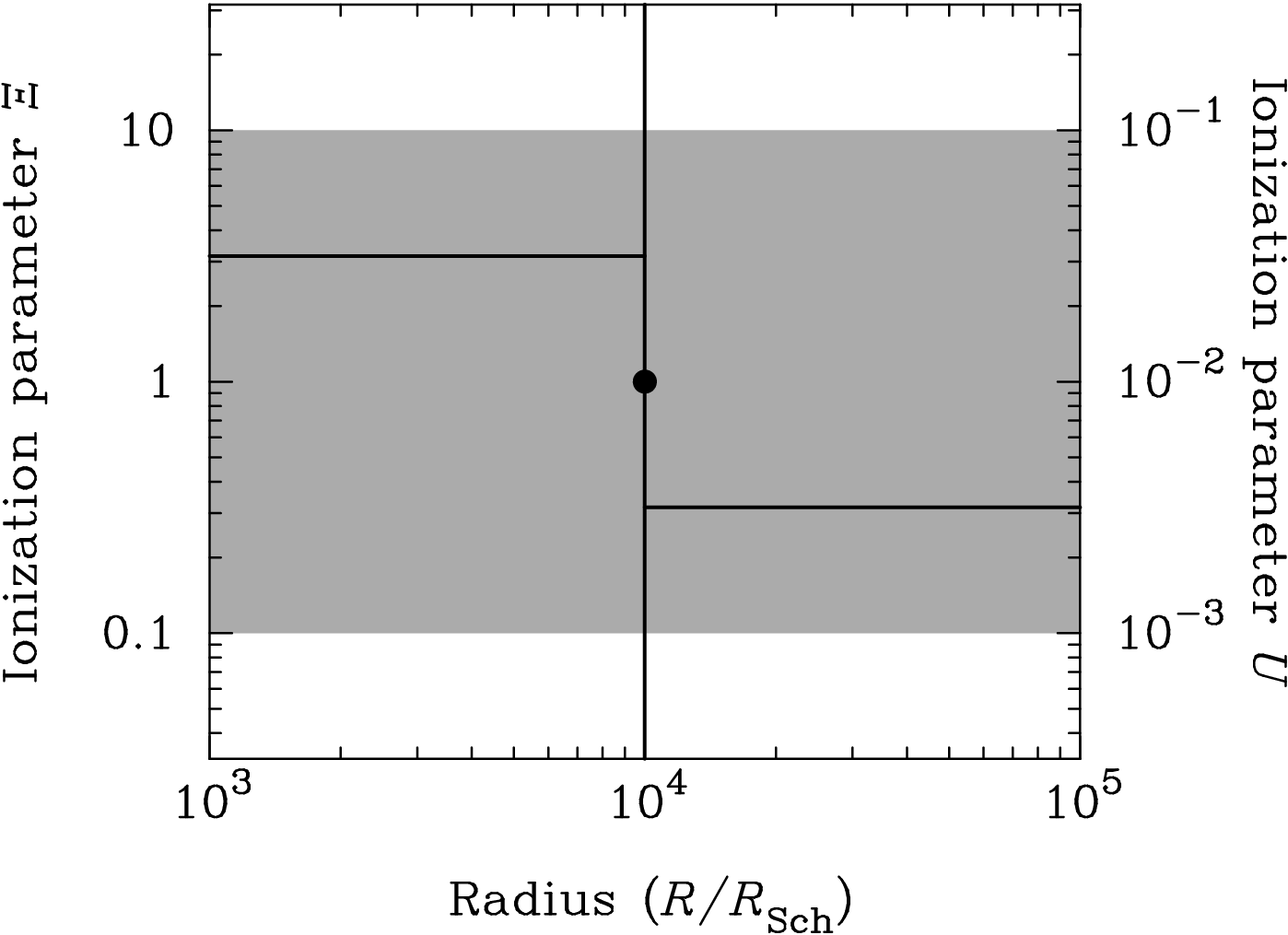}
\figcaption{\footnotesize Ionization parameter $\Xi$. We simply divided broad line regions into
two parts according to the ionization parameter. We use the SED to connect the two different ionization
paramter $U=10^{-2}\Xi$. The shaded part is the possible regime for clouds emitting lines. The $U$ is
usually used to distinguish LIL and HIL regions. The left part is the HIL regions whereas
the right part is the LIL regions. The $\sim 10^4R_{\rm Sch}$ is the boundary for the LIL and HIL regions.
}
\label{phase}
}

\subsubsection{Thermal conduction}
Thermal conduction in the CAS plays a key role in governing the growth of
instability. Its importance can be assessed by considering the heating rate for conduction
${\cal H}_{\rm con}=\nabla\cdot\left(\kappas T^{2.5}\nabla T\right)\sim \kappas T^{3.5}/\delta R^2$,
where $\delta R$ is the scale of the perturbation length of the temperature, and
$\kappas=5.4\times 10^{-7}{\rm erg~s^{-1}~cm^{-1}~K^{-7/2}}$ is
Spitzer's coefficient of thermal conductivity (Spitzer 1962). Comparing it with the free-free cooling
rate gives
${\cal H}_{\rm con}/n^2\calc_{\rm ff}\sim 0.23~T_7^3\delta R_{14}^{-2}n_7^{-2}$,
where $\delta R_{14}=\delta R/10^{14}{\rm cm}$ is the initial length of the thermal perturbation, while
its ratio to the photoionization heating rate is
${\cal H}_{\rm con}/n^2\calh_{\rm ph}\sim 1.0~T_7^{3.9}\delta R_{14}^{-2}n_7^{-2}$, showing that conduction
must be considered in the energy balance.  When the perturbation length $\delta R$
is smaller than $10^{14}$cm, thermal conduction will smear the perturbation within the thermal timescale
and prevent development of the thermal instability and formation of clouds.

As a summary, we have the net heating function
\begin{equation}
\call=\calh_{\rm Comp}+\calh_{\rm ph}-\calc_{\rm Comp}-\calc_{\rm ff}-\calc_{\rm line},
\end{equation}
which are approximated by equations (6-10) for a hot plasma with temperature of $T>10^4\rmK$. It should
be noted that unlike other processes, thermal conduction is excluded in this equation, but it appears in
equation (33) since it only happens between a hot and a cold medium. Thermal
instability of isobaric gas at a constant pressure occurs when $\partial \call/\partial T>0$, since
the heating rate rises faster than the
cooling rate, determining the fate of the CAS that has arisen from the SF disk. Figure \ref{HC} shows the
heating and cooling functions versus temperature at different radii and metallicity. We set $\mbh=10^8\sunm$,
bolometric luminosity $L_{\rm Bol}=10^{46}$\ergs\, and the SED of the photoionizing source is from Figure
\ref{sed_T}.

\subsection{Heated atmosphere: geometry}
The warm skin above the SF disk is inevitably heated by the central engine, producing a new static
equilibrium. The heated and expanded skin is referred to as the Compton atmosphere (CAS) because it
has the Compton temperature, $T_{\rm Comp}\sim10^7$K (see \S2.2.3). Its height is
\begin{equation}
\frac{H_{_{\rm CAS}}}{R}\approx 0.23~ t_{\rm C}^{1/2}r_4^{1/2},
\end{equation}
where $t_{\rm C}=T_{\rm Comp}/3\times 10^7$K. The higher the Compton temperature, the thicker the CAS.
Since the CAS has angular momentum taken from the SF disk, it forms a flattened disk around the axis
perpendicular to the SF disk. As we discuss below, cold clouds are born through the thermal instability
in the CAS, but they do not always follow the dynamics of the CAS. Once the cold clouds are formed from the
CAS, their dynamics and fates are governed by the SMBH gravity, radiation pressure from the central engine
and the friction of the CAS. Depending on their column density and metallicity,
cold clouds are either blown away as cloud outflows driven by the radiation pressure, or sink backward onto
the SF disk, or are destroyed by the dynamical friction of the CAS and then recycle back into the CAS.

We would like to point out that the Compton temperature is determined by the spectral energy distribution
(SED) which depends on the Eddington ratios (Wang et al. 2004). This implies that the BLR geometry might
relate to the Eddington ratios, namely, the properties of broad emission lines correlate with the Eddington
ratios. This is indeed true in large samples of quasars (Marziani et al. 2009), where the eigenvector
1 spectra depends on the Eddington ratio. Finally the mass ratio of the Compton gas converting
into cold clouds remains a free parameter in the present model. A self-consistent model could resolve
this problem.

\subsection{Thermal equilibrium and gas supply}

For a typical luminous QSO with continuum luminosity $L_{1450} = 10^{44}~{\rm erg~s^{-1}~\AA^{-1}}$, the total
mass needed to provide the emission lines seen from the full BLR is of order $10^3\sunm$ (Baldwin et al. 2003a).
This estimate includes neutral material in the interiors of optically thick clouds, and we use it here for the
total mass of the LIL clouds, which we designate $ M_{\rm LIL}$. For the case of the optically-thin HIL clouds,
we use just the zones within the clouds which produce \civ\ emission, which would correspond to
$M_{\rm HIL} = 3\sunm$ for the same continuum luminosity. Following Peterson (1997), we round this down to
$M_{\rm HIL}\sim 1\sunm$.

The timescales for supplying the HIL and LIL masses are given by
\begin{equation}
t_{\rm HIL}=\frac{M_{\rm HIL}}{\dot{M}_{\rm diff}},~~~
t_{\rm LIL}=\frac{M_{\rm LIL}}{\dot{M}_{\rm diff}}.
\end{equation}
We find that $t_{\rm HIL}\sim 10$yr for typical $M_{\rm HIL}\sim 1\sunm$ and $\dot{M}_{\rm diff}\sim 0.1\sunmyr$
($R=10^3\sim 10^4R_{\rm Sch}$), and $t_{\rm LIL}\sim 10^2-10^3$yr for $M_{\rm LIL}\sim 10^2-10^3\sunm$ and
$\dot{M}_{\rm diff}\sim 1\sunmyr$ ($R=10^4\sim 10^5R_{\rm Sch}$).
It generally follows that $t_{\rm HIL}> t_{\rm Comp}$ and $t_{\rm LIL}> t_{\rm Comp}$. This means that
all the supplied gas will be efficiently heated to the Compton temperature until the two timescales are
equal. The CAS keeps a quasi-static equilibrium state so that we use a linear analysis to treat its
thermal instability, which depends on the equilibrium.

\begin{figure*}
\centering
\figurenum{5}
\includegraphics[angle=-90,scale=0.56]{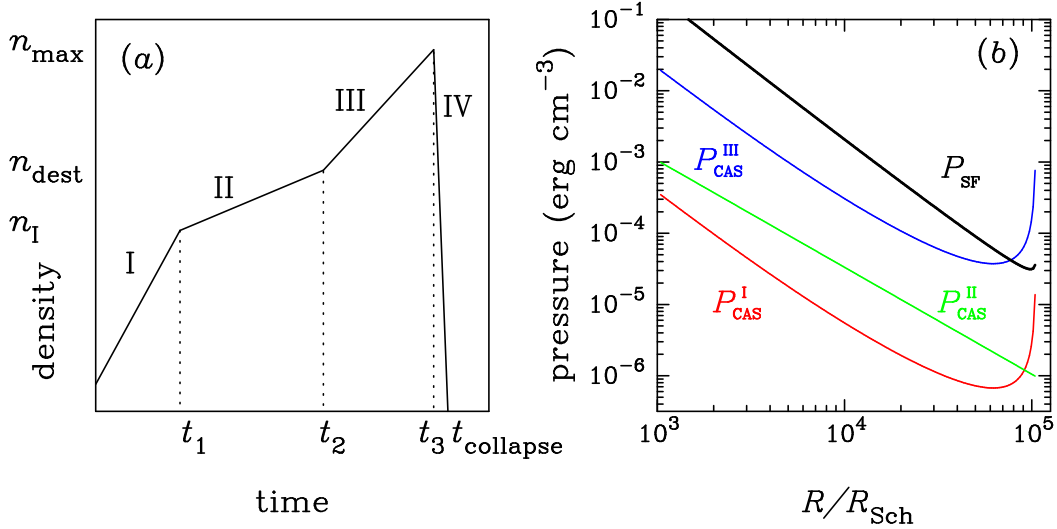}
\figcaption{\footnotesize {\em Left:}
The four phases of the evolution of the Compton atmosphere as it is supplied by the SF disk.
Phase I is a duration of pure accumulation of the diffused gas. Phase II is
a period of formation of cold clouds and some clouds sink to the SF disk. In phase III, clouds are still
born and destroyed through dynamic interaction with the atmosphere, finally giving rise to collapse into the SF
disk in Phase IV. See text for details of the episodic behaviors of the atmosphere.
{\em Right:} Comparison of the thermal pressure in the SF disk with that in phase I, II and III. It is
clear that $P_{\rm SF}$ is always larger than $P_{\rm CAS}^{\rm I}$, $P_{\rm CAS}^{\rm II}$ and
$P_{\rm CAS}^{\rm III}$, giving rise to continuous injection of the warm gas into the BLR. The time
in the {\em left} panel is not shown to scale.
}
\label{phase}
\end{figure*}

\subsection{Fates of the clouds}
Once clouds have formed, they are subject to destruction by various dynamical processes provided the CAS is
sufficiently dense (see Mathews \& Blumenthal 1977; Krolik et al. 1981; Mathews 1986 for details of these destruction
processes). If they are not destroyed instantly, they may undergo one of three different fates depending on
their properties and the CAS density. If the clouds' column density is small enough, radiation pressure may
blow them away from the region where they born, forming outflows (Boroson 2005) and increasing the metallicity in narrow
line regions. This process simply decreases the total mass of the atmosphere. The second possibility is for
the clouds to sink to the SF disk since they are too heavy to be supported by the buoyancy and radiation
pressure (as shown in \S3.3.2). This also simply decreases the total mass of the atmosphere. The third possible
fate is for a cycle between the CAS and the clouds (Krolik 1988). The clouds are destroyed by the dynamic
interaction with the CAS and return mass to the CAS, which then forms new clouds. This opens the way to an
eventual collapse of the CAS. The mass cycle does not decrease the total mass of the atmosphere, while at
the same time the continuous supply from the SF disk steadily increases the total mass of the CAS.
If then at some point the mass stored in the clouds is rapidly released into the atmosphere through
dynamical destruction, the optical depth of the atmosphere is suddenly amplified, making the irradiation
insufficient and the cooling of the CAS very efficient. This induces
the CAS to catastrophically collapse onto the SF disk. This process drives episodic appearance of the BLRs,
at least in some parts. We will discuss these complicated processes in \S4.

Those optically thin clouds formed in the innermost regions are almost instantly blown away by the
radiation pressure. This might be related to the formation of narrow line regions (Wang et al. 2011 in
preparation). Evidence for this is the tight correlation between the BLR and the
NLR metallicities (Wang et al. 2011b). The acceleration of the clouds in this region is discussed in \S4.2.

\section{The Compton atmosphere: the birthplace of cold clouds}

Once the supplied gas is exposed to the central engine, it will be heated up to the Compton temperature
and will become a rotating flattened disk of Compton gas. We stress that the continuous injection into
the atmosphere from the SF disk leads to a complicated evolution of the atmosphere. The CAS is undergoing
heating by the accretion disk and radiative cooling, and also is carrying the Keplerian angular momentum
from the SF disk. The resulting gas flows are reflected in the initial motions of the cold clouds that
form from the CAS, and thus are reflected in the observed profiles of broad emission lines.

\subsection{The accumulation of gas in an evolving Compton atmosphere}
\subsubsection{Mass budget}

We first consider the mass budget of the CAS. For simplicity, we assume the cold clouds which form from
the CAS all have the same mass and size unless we point out the difference in the LIL and HIL.

The CAS evolution can generally be divided into the four phases illustrated by Figure \ref{phase}: phase
I: accumulation of the atmosphere is simply governed by the injection from the SF disk; phase II: formation
and pile-up of clouds are driven by the thermal instability along with significant downward-spiraling of
clouds to the SF disk; phase III: the atmosphere is so dense that sinking clouds are destroyed by
dynamical friction, forming a cycle between clouds and atmosphere, and  phase IV: the dense atmosphere
is not supported by the gas pressure, giving rise to a collapse of the atmosphere into the SF disk.

Phase I continues until the time $t_1$ when the line cooling dominates over the Compton
cooling. During this phase, we have
\begin{equation}
\mcas(t)=\int_0^{t~(\le t_1)} \dot{M}_{\rm inj}dt.
\end{equation}
At $t_1$, the CAS enters phase II. Cold clouds form and separate from the CAS atmosphere.
During phase II, some clouds sink to the SF disk and we have
\begin{equation}
\dot{M}_{_{\rm CAS}}+\dot{N}_{\rm c}m_c=\dot{M}_{\rm inj}
\end{equation}
where $\dot{N}_c$ is the birth rates of clouds, and $m_c$ is the mass of each cloud. Since the CAS
is not yet very dense, the damping can be neglected in this phase. The sinking velocity of clouds can
be estimated by $v_{\rm sink}=v_{\rm K}(\hcas/R)=\hcas\Omega_{\rm K}$ with a timescale
of $t_{\rm sink}=\hcas/v_{\rm sink}=\Omega_{\rm K}^{-1}$, where $v_{\rm K}$ is the Keplerian rotation
velocity. This means that clouds will sink to the SF disk roughly within a Keplerian timescale. It
should be noted that this simple estimation does not include the influence of the CAS friction. The
sinking timescale could be longer then.

The CAS comes into Phase III when it becomes sufficiently dense that dynamical friction destroys the clouds
before they can sink all the way to the SF disk, leading to return of cloud material to the atmosphere. We now have
\begin{equation}
\dot{N}_cm_c+\dot{M}_{_{\rm CAS}}=\dot{M}_{\rm inj}+\dot{N}_{\rm des}m_c,
\end{equation}
and $\dot{M}_{\rm sink}\ll\dot{M}_{\rm des}$, where $\dot{M}_{\rm sink}=\dot{N}_{\rm sink}m_c$. Since
mass continues to be injected from the SF disk, the CAS will become denser and denser until its
temperature balance becomes dominated by line cooling. The catastrophe is then triggered, leading to
a collapse of the CAS onto the SF disk. The collapse timescale ($t_{\rm collapse}$) is determined by
the maximum of the sinking timescale ($t_{\rm sink}$) and cooling timescale ($t_{\rm cool}$), namely,
\begin{equation}
t_{\rm collapse}=\max(t_{\rm sink},t_{\rm cool}).
\end{equation}
In the following sections, we will determine the critical times and the corresponding densities.

There is a necessary condition for continuous injection from the SF disk into the CAS, namely, the
gas pressure in the SF disk ($P_{\rm SF}$) should be higher than that of the CAS ($P_{\rm CAS}$).
Otherwise, the injection will stop if $P_{\rm SF}\le P_{\rm CAS}$. These pressures are compared in
Figure 5{\em b}, and it is seen that $P_{\rm SF} > P_{\rm CAS}$ throughout Phases I-III. This guarantees
the continuous injection to the CAS from the SF disk.

\subsubsection{Critical times and densities}
For a fully ionized Compton gas, the radiative acceleration\footnote{Even for a super-Eddington
accreting SMBH, the radiative luminosity is slightly lower than the Eddington luminosity due to photon
trapping effects (Wang \& Zhou 1999), and the Compton gas is still bound. This could happen in narrow line
Seyfert 1 galaxies with super-Eddington accretion rates.} is $a_{\rm rad}=\elledd g$,
where $g=G\mbh/R^2$ is the gravitational acceleration.
Before line cooling dominates, radiation pressure acting on the ions is mainly through Thompson scattering
by electrons. Provided the radiation luminosity of the SMBH is sub-Eddington, the Compton gas is bound
by the SMBH potential since its thermal temperature is lower than the virial temperature.  The Compton gas will
accumulate until it becomes partially ionized. At that point the radiation pressure will be enhanced by
line absorption, and the partially ionized gas will be blown away if the radiation pressure is strong enough.
There are critical densities of the CAS governing its thermal states in light of the cooling and
heating. We discuss the case with $\mbh=10^8\sunm$ and $L_{\rm bol}=3.0\times 10^{45}\ell_{0.3}M_8$\ergs\,
to illustrate the evolution of the CAS.

During phase I, the CAS receives gas through thermal diffusion from the SF disk and is cooled mainly
by the Compton cooling until the CAS density is high enough so that line cooling dominates in the
formed clouds. During phase I, the largest size of a perturbation cannot exceed the vertical height
of the CAS given by equation (15) for a thermal instability which developes as described below by equation (57).
This gives a lower limit of the CAS density as $n_{\rm min}\sim 10^4{\rm cm^{-3}}$.
However, cold clouds cannot be formed because their collapse timescale will be much longer
than the cooling timescale, and so the thermal perturbation is removed by the dynamics (see equation
59, below). Therefore, pure accumulation without formation of clouds happens in phase I. Accumulation
continues until clouds form, and the CAS enters phase II. This transition from phase I to II is
determined by the time at which line cooling in clouds dominates the Compton cooling. The Compton cooling function
is given by
$\calc_{\rm Comp}\approx 1.8\times 10^{-23}~r_3^{-2}M_8^{-1}\ell_{0.3} t_{\rm C}n_8^{-1}{\rm erg~cm^3~s^{-1}}$.
When the CAS density exceeds $n_{\rm I}$, for a plasma with temperature
$10^5{\rm K}\le T_e\le 10^7$ K, the CAS begins to form clouds. With
$\calc_{\rm line}\approx 10^{-21}Z_0~{\rm erg~cm^3~s^{-1}}$ from Figure \ref{HC}, we have
\begin{equation}
n_{\rm I}=1.8\times 10^6~M_8^{-1}\ell_{0.3}t_{\rm C}Z_0^{-1}r_3^{-2}~{\rm cm^{-3}}.
\end{equation}
The total mass of the CAS is given by the integration over the entire CAS. We have
\begin{equation}
\mcas^{\rm I}=25.3~M_8^2\ell_{0.3}t_{\rm C}^{3/2}Z_0^{-1}r_5^{3/2}~\sunm,
\end{equation}
where $r_5=R/10^5R_{\rm Sch}$ is the outer boundary of the BLR, and the time is roughly given by
\begin{equation}
t_1=\frac{\mcas^{\rm I}}{\dot{M}_{\rm inj}}
   =25.3~M_8^2\ell_{0.3}t_{\rm C}^{3/2}Z_0^{-1}r_5^{3/2}\dot{M}_{\rm inj,0}^{-1}~{\rm yr}.
\end{equation}
where $\dot{M}_{\rm inj,0}=\dot{M}_{\rm inj}/1.0\sunmyr$ is the diffusion rate given by equation (3).
During phase II, once clouds have formed, they will spiral down to the SF disk with a sink velocity
estimated by $v_{\rm sink}=v_{\rm K}(H/R)$, where $v_{\rm K}$ is the Keplerian rotation velocity. This
estimate is based on fact that the cloud motions are mainly controlled by the vertical gravity of the
SMBH. We have the sink timescale $t_{\rm sink}=H/v_{\rm sink}=\Omega_{\rm K}^{-1}$, which is just the
Keplerian one.

There will be a dynamical interaction due to the velocity difference between the clouds and the CAS.
This eventually can destroy the clouds and return their material to the CAS, constituting a mass cycle.
The destruction timescale is given by
$t_{\rm dest}=3.4\times 10^7~\Xi^{-1/2}N_{22}(t_{\rm C}T_4)^{1/4}(n_eT_e)_{14}^{-1}\Delta v_{500}^{-1}$
sec, where $\Delta v_{500}=\Delta v/500\kms$ is the velocity difference (Krolik 1988). Setting
$t_{\rm sink}=t_{\rm dest}$, we have
\begin{equation}
n_{\rm dest}=2.5\times 10^6~\Xi^{-1/2} N_{22}t_{\rm C}^{-3/4}T_4^{1/4}M_8^{-1}\Delta v_{500}^{-1}r_3^{-3/2}
     ~{\rm cm^{-3}},
\end{equation}
where $N_{22}=N_{\rm cl}/10^{22}{\rm cm^{-2}}$ is the column density of clouds. Ferland et al. (2009)
show that the typical column density of \feii clouds is of order $10^{24}{\rm cm^{-2}}$. We then have,
at the point where the cloud destruction takes over and mass cycling begins, the CAS mass
\begin{equation}
\mcas^{\rm II}=2.6\times 10^2~\Xi^{-1/2} N_{22}(T_4/t_{\rm C})^{1/4}\Delta v_{500}^{-1}r_5^2M_8^{2}~\sunm,
\end{equation}
corresponding to the time $t_2$
\begin{equation}
t_2=\frac{\mcas^{\rm II}}{\dot{M}_{\rm inj}}
   =2.6\times 10^2~\Xi^{-1/2} N_{22}(T_4/t_{\rm C})^{1/4}\Delta v_{500}^{-1}r_5^2M_8^{2}\dot{M}_{\rm inj,0}^{-1}~{\rm yr}.
\end{equation}
Since the formation of cold clouds decreases the gas pressure in the CAS, $P_{\rm SF}>P_{\rm CAS}$ drives
the continuous injection into the CAS, entering phase III. However, the start of the mass cycle between
the clouds and the Compton gas in the CAS will cause an increase in the gas pressure in the CAS. The CAS
reaches a steady state with a balance between production and destruction of clouds, entering phase III.

Once mass begins cycling back and forth between the clouds and the CAS, there is no longer a loss of gas back
onto the SF disk, and the CAS density will increase more rapidly in response to the injection from the SF disk.
This increase will eventually cause the CAS to enter phase IV, where rather than Compton cooling the entire CAS
is cooling mainly through emission lines from H-like Fe ions. In such a case, the cooling function is
approximated by $\calc_{\rm Fe^{+25}}=2.0\times 10^{-23}~Z_0~ {\rm erg~cm^3~s^{-1}}$,
where $t_{\rm C}=T/T_{\rm Comp}$ and $T_{\rm Comp}=3\times 10^7$K. Setting $\calc_{\rm Comp}=\calc_{\rm line}$,
we have as the critical density for the start of Phase IV (the collapse phase)
\begin{equation}
n_{\rm max}\approx 1.0\times 10^8~M_8^{-1}\ell_{0.3} t_{\rm C}Z_0^{-1}r_3^{-2}~{\rm cm^{-3}},
\end{equation}
with a scale height of $\hcas$ given by equation (15). The corresponding mass is roughly obtained through
integrating the regions with the density profile (equation 27), giving
\begin{equation}
\mcas^{\rm III}\approx 1.3\times 10^3~M_8^2\ell_{0.3}t_{\rm C}^{3/2}Z_0^{-1}r_5^{3/2}~\sunm,
\end{equation}
and the time is roughly
\begin{equation}
t_3=\frac{\mcas^{\rm III}}{\dot{M}_{\rm inj}}
   =1.3\times 10^3~M_8^2\ell_{0.3}t_{\rm C}^{3/2}Z_0^{-1}\dot{M}_{\rm inj,0}^{-1}r_5^{3/2}~{\rm yr}.
\end{equation}
Since the entire CAS is cooling with a timescale (see equation 62) much shorter than the injection timescale,
a direct collapse of the CAS including the clouds is inevitable within timescale
\begin{equation}
t_{\rm collapse}\approx t_{\rm sink}\approx \frac{R}{v_{\rm K}}=45.1~M_8r_4^{3/2}~{\rm yr}.
\end{equation}
Following such a collapse, the continued injection from the SF disk drives the BLR to undergo episodic
reappearances during the AGN lifetime.

In Figure \ref{phase}{\em b} we compare the SF disk pressure with the CAS pressure during phases I-III.
We use $(\pcas^{\rm I},\pcas^{\rm II},\pcas^{\rm III})=(n_{\rm I},n_{\rm dest},n_{\rm max})kT_{\rm Comp}$
for the three phases, respectively. The metallicity appearing in the critical density is given by the
results in W11. We find that the SF disk pressure is always higher than the CAS pressure. This simply
indicates that the SF disk continuously injects material into the CAS. This is a necessary condition
for the episodic appearance of the BLR. On the other hand, the injection timescale is much longer than
the dynamical timescale, allowing us to approximate the CAS by a static state at any time except during
Phase IV.

It is interesting to find that the accumulated mass of the BLR agrees well with the observational estimates
given above in \S 2.4. Our model predicts that BLR-I (the HIL gas) and BLR-II (the LIL gas) are spatially
separated, and have $\sim 1\sunm$ and $\sim 300\sunm$, respectively. The estimates based on the observed
emission line strengths are of order $M_{\rm HIL}\sim 1\sunm$ and $M_{\rm LIL}\sim 10^3\sunm$. This
provides independent support for the present model of the BLR formation.

We find that the maximum CAS mass is very sensitive to the SMBH mass, but also to the Eddington ratio,
implying the dependence of the BLR mass on the SMBH mass and accretion rate through the star formation rate.
For narrow line Seyfert 1 galaxies with $\mbh\sim 10^{6.5}\sunm$, the relevant timescales could be
significantly reduced, but relying on accretion rates. We will consider the case of narrow
line Seyfert 1 galaxies in the future.

It not expected that significant soft X-rays are emitted from the CAS. The emission from free-free
cooling is estimated to be $L_{\rm SX}=4\pi \int n_e^2\calc_{\rm ff}RH_{\rm CAS}dR$. The phase I CAS emits
$L_{\rm SX}\sim 2.0\times 10^{39}$\ergs\, and the Thompson scattering depth is $\tau_{\rm es}\approx 0.4$.
This soft X-ray emission can be totally neglected compared with the bolometric luminosity. We should point
out that most of the injected mass $\mcas$ will be rapidly converted into clouds after phase I, but the CAS
gas will emit soft X-rays during the formation of clouds. In phase II, the injected mass is efficiently
converted into clouds, and the luminosity from the CAS cooling is given by
$L_{\rm SX}\sim kT_{\rm C}\dot{M}_{\rm inj}/m_p\approx 1.7\times 10^{41}$\ergs. This again is not
significant.

We note that the continuous injection could be stopped if pressure of the Compton atmosphere is higher than
the pressure of the hot gas in the star forming disk. This could happen if the star formation rates are
not high enough in the case of some low luminosity AGNs (LLAGNs). Also thermal conduction between the CAS
and the warm gas in the SF disk could be important. The underlying physics will be discussed separately for
the LLAGNs. We should point out here that the above four phases are just an outline of the complicated
evolution of the atmosphere. The four phases of the atmosphere are radius-dependent, forming the low-
ionization-line regions and high-ionization-line regions at different times. We will investigate the four
phases in further detail in the following sections.

\subsection{Thermal instability}
It has been long known that the partially ionized gas exposed to a radiation field can undergo thermal
instability (Field 1965). It develops when $\partial \call/\partial T<0$.
Formation of BLRs driven by the thermal instability has
been extensively studied (Krolik et al. 1981; Beltrametti 1981; Shlosmann et al. 1985; Krolik 1988; or
more recently Pittard et al. 2003 for the more complicated case of winds). Basic results from these studies
show that a two-phase medium will be formed through the thermal instability. However, most of
results do not directly apply to the present case because the authors did not include the role of the
angular momentum of the gas that is undergoing the thermal instability. This plays a key role in determining the geometry of the
resulting BLR. Furthermore all these studies neglect the role of metallicity in the condensation of the clouds.
Beltrametti (1981) demonstrated that cold clouds are formed at a 1 pc scale for metal-free
gas, clearly larger than $R_{\rm BLR}\approx 0.1 L_{46}^{1/2}$pc obtained from reverberation mapping,
where $L_{46}=L_{\rm Bol}/10^{46}~{\rm erg~s^{-1}}$ is the bolometric luminosity. We argue that the
inconsistency is due to the metal-free assumption (Beltrametti 1981; Shlosman et al. 1985). A detailed
study of the thermal instability of the atmosphere including the angular momentum and metallicity is
one of main goals of the present paper. We start from the basic equations of the atmosphere.

\subsubsection{Governing equations}
The atmosphere is described by a velocity field $\vec{v}$, density $\rho$, temperature $T$, and pressure
$p$, which are a function of radius, in the potential ($\Phi$) of the SMBH gravity. The SF disk as the
lower boundary of the atmosphere continuously injects warm gas with a function $\calS$.
Thermal conduction in the atmosphere is included in the energy equation. The atmosphere can be descried
by a series of equations: 1) the continuity equation
\begin{equation}
\frac{\prho}{\pt}+\nabla\cdot\left(\rho \vec{v}\right)=0;
\end{equation}
2) momentum conservation
\begin{equation}
\frac{\pp \vec{v}}{\pt}+\left(\vec{v}\cdot\nabla\right)\vec{v}=-\frac{1}{\rho}\nabla p-\nabla\Phi;
\end{equation}
and 3) energy conservation
\begin{equation}
\begin{array}{ll}
\displaystyle\frac{p}{\gamma-1}\left(\frac{\pp}{\pt}+\vec{v}\cdot\nabla\right)\ln\left(p\rho^{-\gamma}\right)&=
\nabla\cdot\left(\kappas T^{2.5}\nabla T\right)\\
&\quad+\displaystyle\left(\frac{\rho}{\mu m_p}\right)^2\call(\rho,T),
\end{array}
\end{equation}
where $\call(\rho,T)=\calh-\calc$ is the net heating function, $\calh$ and $\calc$ are heating and
cooling functions, respectively, and here $\gamma$ is the ratio of principle specific heats of the
gas. We assume the system has a cylindrical asymmetry so
that all terms of $\partial /\partial \phi=0$. In cylindrical coordinates, the above equations can
be re-cast in three components
\begin{equation}
\frac{\prho}{\pt}+\frac{1}{R}\frac{\pp}{\pr}\left(R\rho\vr\right)+
\frac{\pp}{\pz}\left(\rho \vz\right)=0,
\end{equation}
\begin{equation}
\frac{\pp\vr}{\pt}+\vr\frac{\pp\vr}{\pr}
+\vz\frac{\pp\vr}{\pz}=
-\frac{1}{\rho}\frac{\pp p}{\pr}-\frac{\pp\Phi}{\pr}+\frac{\vphi^2}{R},
\end{equation}
\begin{equation}
\frac{\pp\vz}{\pt}+\vr\frac{\pp\vz}{\pr}
+\vz\frac{\pp\vz}{\pz}=
-\frac{1}{\rho}\frac{\pp p}{\pz}-\frac{\pp\Phi}{\pz},
\end{equation}
\begin{equation}
\frac{\pp\vphi}{\pt}+\vr\frac{\pp\vphi}{\pr}
+\vz\frac{\pp\vphi}{\pz}=
-\frac{\vr\vphi}{R}.
\end{equation}
\begin{equation}
\begin{array}{ll}\displaystyle
\frac{p}{\gamma-1}&\!\displaystyle\left[\frac{\pp}{\pt}+\vr\frac{\pp}{\pr}+\vz\frac{\pp}{\pz}
\right]\ln\left(p\rho^{-\gamma}\right)=\displaystyle \frac{1}{R}\frac{\pp}{\pr}
          \left(R\kappas T^{5/2}\frac{\pp T}{\pr}\right) \\
&\qquad\qquad\quad+\displaystyle\frac{\pp}{\pz}\left(\kappas T^{5/2}\frac{\pp T}{\pz}\right)
+\left(\frac{\rho}{\mu m_p}\right)^2\call(\rho,T).
\end{array}
\end{equation}
The potential of the SMBH gravity is given by
\begin{equation}
\Phi(R,z)=-\frac{G\mbh}{(R^2+z^2)^{1/2}},
\end{equation}
where $z$ is height from the SF disk plane. Here the 3-dimensional velocity components are
$\vr$, $\vphi$ and $\vz$.

The dynamical timescale of the CAS gas is much shorter than the timescale of gas supply. This
allows us to assume that the gas maintains hydrostatic and thermal equilibrium at any time and treat the gas
supply as a small perturbation. In order to proceed to study the global instability of the atmosphere, we
simply assume that the injection function $\calS\approx 0$ and obtain the solution for the barotropic gas.
We emphasize that the injection is so important for the BLR that it may drive a transient appearance
of a BLR in some AGNs.

\begin{figure*}
\centering
\figurenum{6}
\centerline{\includegraphics[angle=-90,scale=0.7]{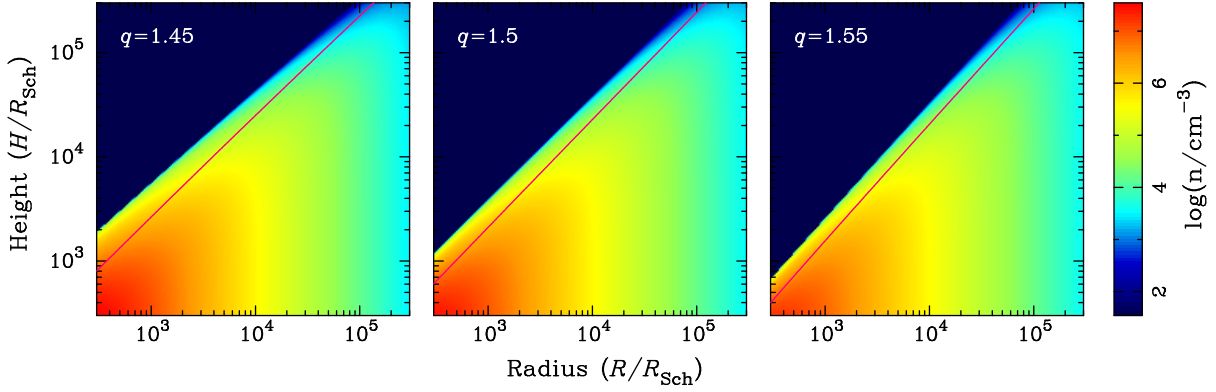}}
\figcaption{\footnotesize
The iso-density profiles of the barotropic model of the atmosphere. We use $n=1.4$, $\omega_0=0.5$
and $\rho_{*0}=5.8\times 10^{-20}{\rm g~cm^{-3}}$, but change the index of angular momentum ($q$),
and the constant $C_0=0$ for each case. The total mass of the CAS is $\mcas=1.3\times 10^3\sunm$.
The red line is the boundary of the atmosphere beyond which clouds are not able to form.}
\label{density}
\end{figure*}

\subsubsection{Equilibrium configurations}
The detailed thermal behavior of the atmosphere is sensitive to the gas temperature and density distribution,
so the conclusions drawn in the cases of clusters and galaxies cannot be a priori extended to the BLR atmosphere.
The equilibrium condition of the gas is important not only because the thermal instability depends on the
equilibrium configuration, but also because the clouds formed due to the instability follow the density
distribution in the equilibrium state. When the Compton cooling and heating reach equilibrium, the gas becomes
a Compton gas with a uniform temperature. We give the equilibrium states of the barotropic gas.
For a static state of the atmosphere, $\vr=\vz=0$.

The Poincar\'e-Wavre theorem states that the surfaces of constant pressure and constant density
coincide if and only if $\pp\Omega/\pp z=0$. Although the assumption of a barotropic gas is simplified, it
illustrates the main features as a good approximation. The rotating atmosphere in cylindrical coordinates
can be described by the following equations re-cast from (34)-(37)
\begin{equation}
\frac{1}{\rho_0}\frac{\pp p_0}{\pr}+\frac{\pp\Phi}{\pr}-\frac{\vphi^2}{R}=0,
\end{equation}
\begin{equation}
\frac{1}{\rho_0}\frac{\pp p_0}{\pz}+\frac{\pp\Phi}{\pz}=0.
\end{equation}
For a barotropic gas, it follows that $p_0=K_0\rho_0^{\gamma}$ and the angular velocity $\Omega$ is a
function only of $R$. The analytical solution of equations (40) and (41) is given by
$\gamma p_0/(\gamma-1)\rho_0+\Phi=\int\Omega^2R dR$ for $\gamma\neq 1$.
For a presumed rotation of the atmosphere as
\begin{equation}
\Omega=\omega_0\Omega_0\left(\frac{R}{\rout}\right)^{-q},
\end{equation}
where $\Omega_0=\left(G\mbh/R_{\rm out}^3\right)^{1/2}$ is a constant, and $\omega_0$ is a constant for
correction of the angular momentum, we have the equilibrium solution
\begin{equation}
\frac{p_0}{\rho_0}=\frac{\gamma-1}{\gamma}\left[C_0-\Phi+\frac{\omega_0^2\Omega_0^2}{2-2q}
            \left(\frac{R}{R_{\rm out}}\right)^{-2q}R^2\right],
\end{equation}
where $C_0$ is a constant to be determined by the total mass of the atmosphere. Appendix A gives the solutions
for $\gamma=1$. The zero-pressure surface is given by
\begin{equation}
\Phi-\frac{\omega_0^2\Omega_0^2}{2-2q}\left(\frac{R}{R_{\rm out}}\right)^{-2q}R^2=C_0.
\end{equation}
It is convenient to take the constant $\Omega_0=c/\sqrt{2}R_{\rm Sch}r_{\rm out}^{3/2}$, which is the
Keplerian rotation velocity. Equation (44) then can be re-cast as
\begin{equation}
\frac{\omega_0^2r_{\rm out}^{2q-3}}{2(q-1)}\frac{1}{r^{2(q-1)}}-\frac{1}{(r^2+h^2)^{1/2}}=C_0^{\prime},
\end{equation}
where $r=R/R_{\rm Sch}$, $h=z/R_{\rm Sch}$ and $C_0^{\prime}=2C_0/c^2$. The properties of the
zero-pressure surface are fully determined by the index $q$ and $C_0^{\prime}$. The case of a constant
specific angular momentum has been discussed by Papaloizou \& Pringle (1984). In the present case, we
assume that the atmosphere
retains the Keplerian angular momentum carried from the SF disk. Actually, the atmosphere will
re-redistribute the angular momentum through viscosity leading to a change the density distribution. This
redistribution could be a self-regulated process since the total mass could be a few $10^3 \sunm$ injected
from the SF disk with a mass rate of $\sim 1\sunmyr$. The main goal of the present paper is to
explore the solution of the atmosphere with a presumed angular momentum distribution, in order to obtain
information about cloud formation in the broad line regions. A self-consistent solution for the dynamics of the
atmosphere will be carried out in a separate paper.

Since the inner radius of the SF disk is set at the self-gravitating radius, the atmosphere is assumed to
be confined within the radii between $R_{\rm SG}$ and $R_{\rm out}$. For the barotropic gas, the iso-density and zero-pressure
surfaces overlap. Employing the relation of $p_0=K_0\rho_0^{\gamma}$, we have the density
distribution
\begin{equation}
\rho_0(r,h)=\displaystyle \rho_*\left[\frac{\omega_0^2r_{\rm out}^{2q-3}}{4(1-q)}\frac{1}{r^{2(q-1)}}
            +\frac{1}{2(r^2+h^2)^{1/2}}+C_0\right]^n,
\end{equation}
where we take $K_0=kT_{\rm Comp}\rho_{*0}^{1-\gamma}/m_p$ for convenience,
$\rho_*=\rho_{*0}[\gamma/(\gamma-1)]^{-1/(\gamma-1)}(c_s/c)^{2/(1-\gamma)}$, $\rho_{*0}$ is a
constant, and the constant $C_0$ is then given by the total mass of the atmosphere
\begin{equation}
M_{_{\rm CAS}}=\int\int\rho_0(R,z)2\pi RdR dz.
\end{equation}
Here $M_{\rm CAS}$ is the total mass supplied by the thermal diffusion, which is a time-dependent
parameter in the system.

Figure \ref{density} shows solutions of the barotropic models for different distributions of angular
momentum. These results are found to be very sensitive to the distribution of angular momentum in the
atmosphere (equation 42). We concentrate on the sub-Keplerian cases since the
diffusing gas only carries the Keplerian angular momentum, unless there is a mechanism to drive the
atmosphere to rotate with super-Keplerian velocity. For the atmosphere with a sub-Keplerian rotation,
the atmosphere shrinks from its initial size (i.e. $R < R_{\rm out}$), increasing the density of the
atmosphere. The geometry of the atmosphere is actually quite thick, $H/R\sim 1$, but getting thinner
with increasing $q$. This indicates that the BLR would be geometrically thick. The atmosphere has
a sharp upper boundary as shown by Figure 6. The radial distribution of the density
decreases with radius. The rough estimate given by
equation (15) is consistent with the barotropic model. However, the barotropic approximation
should be revisted in future work on the atmosphere. The effect of including redistribution of angular momentum through
viscosity (and even heating by the atmosphere) is worth exploring.

\subsubsection{Perturbation equations and dispersion relation}
We assume that the equilibrium gas is axisymmetric so that all of the terms $\pp/\pp\phi=0$,
and that the fluid rotates differentially with $\Omega=\Omega(R,z)$. The velocity of the gas with density
$(\rho_0)$ is given by
$(\vr^0,\vphi^0,\vz^0)$ for hydrostatic and thermal equilibrium. The velocity perturbations are
$(\vr^1,\vphi^1,\vz^1)$ and the density is $\rho_1$. Supposing an axisymmetric perturbation
happens in a form of $F=F_0+F_1\exp(-i\omega t+i\kr R+i\kz z)$, where $F_0$ is the parameter value at the
equilibrium and the perturbation $|F_1|\ll F_0$, the linearized equations can be written as
\begin{equation}
ik_R\rho_0\vr^{1}+ik_z\rho_0\vz^{1}-i\hat{\omega}\rho_1=0,
\end{equation}
\begin{equation}
-i\hat{\omega}\rho_0\vr^1-2\Omega\rho_0\vphi^1-\cala_{pR}c_0^2\rho_1+ik_Rp_1=0,
\end{equation}
\begin{equation}
-i\hat{\omega}\rho_0\vz^1-\cala_{pz}c_0^2+ik_zp_1=0,
\end{equation}
\begin{equation}
(\Omega+\Omega_R)\vr^1+\omega_z\vz^1-i\hat{\omega}\vphi^1=0,
\end{equation}
and
\begin{equation}
\begin{array}{ll}
(\cala_{pR}-\gamma\cala_{\rho R})\vr^1+&\!\!\!\!(\cala_{pz}-\gamma\cala_{\rho z})\vz^1 \\
&+(i\hat{\omega}-\omega_{\rm d})\gamma\rho_0^{-1}\rho_1-i\hat{\omega}p_0^{-1}p_1=0,
\end{array}
\end{equation}
where $\hat{\omega}=\omega-(\kr\vr^0+\kz\vz^0)$,
$\omega_{\rm d}=\omega_c+\omega_{\rm th}$, $\Omega_R=\pp (\Omega R)/\pp R$,
\begin{equation}
\omega_c=\frac{\gamma-1}{\gamma}\frac{\kappa (\kr^2+\kz^2)T_0^{7/2}}{p_0},
\end{equation}
and
\begin{equation}
\omega_{\rm th}=\frac{\gamma-1}{\gamma}\left(\frac{\rho_0}{\mu m_p}\right)^2\frac{\xi_0\call_0}{p_0},
\end{equation}
where $\call_0=\call(\rho_0,T_0)$ and $\xi_0=\pp\ln \call_0/\ln \rho_0-\pp\ln\call_0/\pp\ln T_0+2$ at
the equilibrium configuration given by equation (43).

We use the approximation of short-wavelength and low-frequency perturbations, with $k_1R\sim k_2R\gg1$.
This approximation is valid in cases where the size of cold clouds is much smaller than the coordinate
$R$, meaning that $Rk_1\gg1$ and $zk_2\gg1$. The case of $Rk_1\gg 1$ is guaranteed since the BLR radius
is much larger than the size of the clouds. However, $zk_2\gg 1$ indicates that the vertical regions
should be much larger than the size of clouds, implying that the approximation is only valid for the
quite geometrically thick disk.

The Appendix gives details of the derivation of the dispersion relation. For barotropic gas, we have the
dispersion relation given by Equation (B27)
\begin{equation}
(\hatn+\omega_{\rm d})(\hatn^2+\omega_{\rm rot}^2)=0.
\end{equation}
We have
\begin{equation}
\hatn=-\omega_{\rm c}-\omega_{\rm th}~~~~{\rm or}~~~~\hatn^2=-\omega_{\rm rot}^2.
\end{equation}
For a Keplerian rotation, we have $\ell\propto R^{1/2}$, we always have $\omega_{\rm rot}>0$, namely,
$\hatn^2<0$, implying that the system is stable. Only if $\omega_{\rm d}<0$ does thermal instability develop, yielding
\begin{equation}
\kappa_{\rm S}\left(\kr^2+\kz^2\right)T_0^{7/2}+\call_0\xi_0\left(\frac{\rho_0}{\mu m_p}\right)^2<0.
\end{equation}
This inequality determines the regions of the thermal instability. Given the equilibrium configuration,
the minimum wavelength of the perturbation can be found from equation (57). Perturbations shorter than the
minimum length will be removed by thermal conduction and no instability can be developed. Since
the perturbation length is much less than the disk height scale, we consider the simple case of
$\kr\sim \kz=2\pi/\lambda_{\rm TI}$, namely, spherical clouds where $\lambda_{\rm TI}$ is the size
of the perturbation which is not able to be smeared by the thermal conduction. We have
\begin{equation}
\lambda_{\rm TI}\ge \frac{2\sqrt{2}\pi \kappas^{1/2}T_0^{7/4}}{|\mathscr{L}_0\xi_0|^{1/2}}
                \left(\frac{\mu m_p}{\rho_0}\right)
                \approx 1.7\times 10^{14}~T_7^{2.4}n_6^{-1}Z_0^{-1/2}~{\rm cm},
\end{equation}
where $\xi_0=2.78$, $\mu=0.67$ and $\call_0=7.77\times 10^{-22}Z_0~{\rm erg~cm^3~s^{-1}}$ for typical
parameter values $T_0=10^7T_7$K and $\rho_0=1.67\times 10^{-18}n_6~{\rm g~cm^{-3}}$. The dependence on
metallicity originates from the line
cooling function. For the line-cooling dominated case, we roughly have $\mathscr{L}\propto T^{-1.3}Z$ and
$\xi_0$ is not sensitive to the temperature and density.
The dependence on temperature results from the thermal conduction and the line cooling function.
The size of the clouds depends on the thermal conductivity $\kappas$, which determines the minimum cloud
size. For the Compton atmosphere, the temperature is nearly homogeneous in space. On the other hand,
the maximum wavelength set by $\lambda_{\rm max}$ is so long that sound waves cannot cross it in a cooling
time, so that compression will not follow cooling and growth is suppressed (Shlosman et al. 1985). This
yields
\begin{equation}
\lambda_{\rm max}\le c_st_{\rm cool}^{\rm ff}\approx 5.2\times 10^{15}T_7n_6^{-1}~{\rm cm},
\end{equation}
where $t_{\rm cool}^{\rm ff}$ is the free-free cooling timescale and $c_s$ is the sound speed of the hot phase.
The column density of the clouds will be $N_{\rm H}=n_e\lambda_{\rm TI}$,
\begin{equation}
N_{\rm H}^0=\frac{2\sqrt{2}\pi \mu\kappas^{1/2}T_0^{7/4}}{|\mathscr{L}_0\xi_0|^{1/2}}
         \approx 1.7\times 10^{20}~T_7^{2.4}Z_0^{-1/2}~{\rm cm^{-2}}.
\end{equation}
It should be noted that this is the initial column density of clouds. Once clouds form, they
efficiently shrink until they reach a new thermal equilibrium with a temperature ($\tcl$) and get a new
column density.

Note that for a CAS with a density lower than $10^4{\rm cm^{-3}}$, the minimum
length of the perturbation will be larger than the height of the CAS. This means that the thermal
instability is not able to develop and formation of clouds is forbidden. This gives the lower limit
of the CAS density for cloud formation.

\subsection{Formation of clouds}
\subsubsection{Initial phase of formation}
Once thermal instability is triggered, the perturbed gas will undergo contraction until clouds finally
form and then maintain a pressure balance with their surrounding medium. These processes are very complicated
(see the review by Meerson 1996; or recent papers of Iwasaki \& Tsuribe 2008; 2009, but which only deal with a
one-dimensional medium). Krolik (1988) investigated cloud formation in an inflow and outflow under the condition
of broken radiative equilibrium, where the convergence or divergence of the inflows and outflows enhance or diminish
the cooling in clouds, giving rise to only algebraic growth of clouds rather than exponential. The present CAS is in
static equilibrium, so the context of cloud formation is different
from the case described by Krolik (1988). Clouds are born swirling around in the CAS. The initial phase of contraction
is driven by the free-free cooling on a timescale of
\begin{equation}
t_{\rm ff}\approx \left\{\begin{array}{l}
          0.58~n_7^{-1}T_7^{1/2}~ {\rm yr}~~(R=10^3R_{\rm Sch}),\\
                                        \\
          5.8~n_6^{-1}T_7^{1/2}~{\rm yr}~~(R=10^4R_{\rm Sch}).
          \end{array}\right.
\end{equation}
We find that this timescale is still smaller than the Keplerian one,
$t_{\rm Kep}= 2\pi/\Omega_{\rm K}\approx 9.0~r_3^{3/2}M_8=283.0~r_4^{3/2}M_8$ yr. Due to the high metallicity, the
subsequent cooling through line emission is much more efficient than bremsstrahlung cooling.

\subsubsection{Final states of cold clouds}
When the density of the Compton gas exceeds $n_{c}^{\rm ff}$ after the initial phase, for a plasma with temperature
$10^4{\rm K}\le T_e\le 10^7$ K, the cooling is mainly through elements C, He, O, Mg and Fe. The cooling
timescale is given by $t_{\rm cool}\approx n_ekT_e/\Lambda_{\rm cool}(n_e,T_e,Z)$, where
the cooling function is approximated by
$\Lambda(n_e,T_e,Z)=n^2\calc_{\rm line}=2.0\times 10^{-10}n_6^2T_6^{-1.3}Z_0~ {\rm ergs^{-1}cm^{-3}}$
(see equation 25 in W11). We then have as the formation timescale of cold clouds
\begin{equation}
t_{\rm cloud}\sim t_{\rm cool}\approx \left\{\begin{array}{l}
 0.05~n_7^{-1}T_7^{2.3}Z_1^{-1}~{\rm yr}~~(R=10^3R_{\rm Sch}),\\
                                         \\
 5.0~n_6^{-1}T_7^{2.3}Z_0^{-1}~{\rm yr}~~(R=10^4R_{\rm Sch}),
 \end{array}\right.
\end{equation}
which is much shorter than the diffusion time of the Compton gas and the Keplerian rotation period.

Following onset of the thermal instability, cold clouds form in the timescale given by equation (62),
within one Keplerian rotation. This allows us to reasonably assume that the cold clouds have
the same angular momentum as gas diffused from the SF disk since the specific angular momentum
of the gas approximately remains constant (neglecting the radiation pressure for the Compton gas).

The temperature of cold clouds can be determined from the new thermal equilibrium condition
$\calh_{\rm Comp}+\calh_{\rm ph}=\calc_{\rm ff}+\calc_{\rm Comp}+\calc_{\rm line}(\Xi)$, while
density can be determined from the pressure equilibrium with the CAS as given by
$\ncl\tcl=\ncas\tcas$, yielding $\ncl=\left(\tcas/\tcl\right)\ncas$ and $\tcl\sim 10^4$K. The
initial mass of a single cloud is given by $\Delta m_c=4\pi \lambda_{\rm TI}^3\ncas/3$. Considering mass
conservation of clouds, we have $\lambda_{\rm cl}^3\ncl=\lambda_{\rm TI}^3\ncas$, where $\lambda_{\rm cl}$
and $n_{\rm cl}$ are the final radius and density of formed clouds. Since pressure equilibrium holds,
we have $n_{\rm cl}T_{\rm cl}=\ncas \tcas$. The final column density of cold clouds is obtained
from
\begin{equation}
N_{\rm H}=\left(\frac{\tcas}{\tcl}\right)^{2/3}N_{\rm H}^0
          \approx 1.7\times 10^{22}~T_7^{3.1}T_4^{-2/3}Z_0^{-0.5}~{\rm cm^{-2}}.
\end{equation}
where $T_4=T_{\rm cl}/10^4$K.
We note that the final temperature depends on metallicity as shown in Figure \ref{HC}.
It is likely that clouds born in the outer BLR are different from ones formed in the innermost
regions of the BLR. Due to the metallicity gradient, this naturally causes the difference between  high- and
low-ionization regions. The evolution of clouds depends on the metallicity of the clouds.

Here we neglect the dependence of the line cooling and photoionization heating on the ionization parameter.
Since the cooling is strongly dependent on the metallicity, the final column density of clouds will be different
from that given in equation (63) because the new thermal equilibrium temperature for $10Z_\odot$ will be of
$10^5$K (from Figure \ref{HC}). Future papers in this series will treat the clouds in a more self-consistent
manner using photoionization models, and will study the dependence of cloud properties on
the metallicity.

\subsection{Global structure of the BLR}
\subsubsection{Spatial distribution of clouds}
The CAS density profile is described by equation (46), which in turn determines the spatial distribution
of the cloud's formation rate in the torus since $\delta \rho\propto \rho_0$, where $\delta \rho$ is the
overdensity of the cloud compared with its surroundings. The spatial distribution of cold clouds generally
follows the hot medium, but with the constraint that clouds can only form in locations where the cooling
timescale is shorter than the sound crossing timescale. Furthermore, for an initial perturbation with
length $\Delta R$, a necessary condition for cloud formation is
$\Delta R_{\rm min}=1.7\times 10^{14}~T_7^{2.4}n_6^{-1}Z_0^{-1/2}$cm should be
less than the local height of the CAS. We write this condition as $\Delta R_{\rm min}=\alpha_{\rm cl} \hcas$,
where $\alpha_{\rm cl}$ is a (poorly known) fraction of the CAS height below which clouds can form. We have a
critical density for cloud formation
\begin{equation}
n_{\rm min}\ge 4.2\times 10^4~\alpha_{\rm cl,0.1}^{-1}M_8^{-1}T_7^{1.9}r_4^{-3/2}Z_0^{-1/2},
\end{equation}
where $\alpha_{\rm cl,0.1}=\alpha_{\rm cl}/0.1$. The parameter
$\alpha_{\rm cl}^{-1}\sim \hcas/\Delta R_{\rm cl}$ determines the number of clouds along the vertical
direction in the CAS, where $\Delta R_{\rm cl}$ is the size of the formed clouds. Clouds can only form
where the density is greater than $n_{\rm min}$. The iso-density plot in Figure \ref{density} shows the
geometry of the BLR with the CAS density indicated by the color. Once clouds
have formed, their fate will depend on their column density.

\begin{figure*}
\centering
\figurenum{7}
\centerline{\includegraphics[angle=-90,scale=0.7]{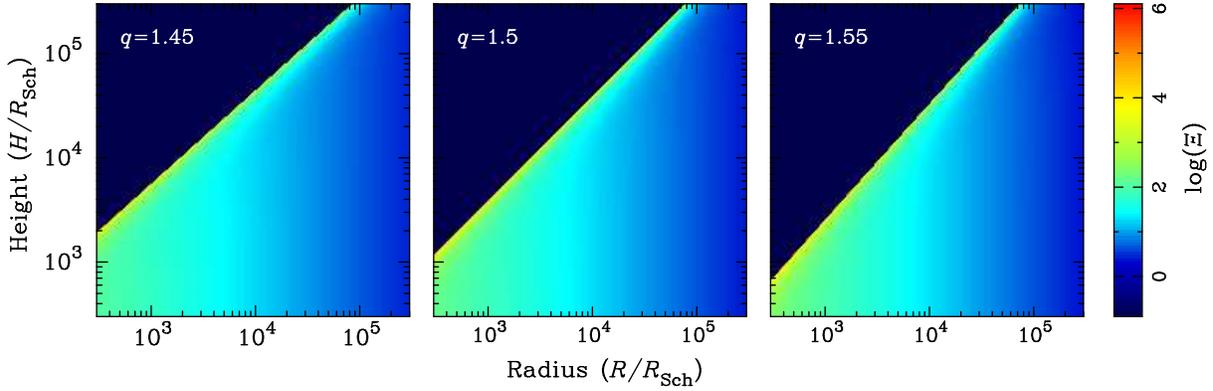}}
\figcaption{\footnotesize
Ionization parameters in the CAS. The three panels correspond to the density profiles shown in Figure \ref{density}.
We use $L_{\rm ion}=10^{45}$\ergs.
}
\label{ionization}
\end{figure*}

\subsubsection{Dynamics of clouds}
Following their formation, the clouds are decoupled from the atmosphere and move in response to the gas
pressure gradient, the radiation pressure and the SMBH gravitational potential.  The vertical support
from the gas pressure gradient can be estimated by
$F_{\rm gas}\sim (\Delta R_c/H_{\rm CAS})P_{\rm gas}\Delta R^2= \rho_{_{\rm CAS}}kT_{\rm Comp} (\Delta R_c)^2/m_p
            =\rho_{_{\rm CAS}}T_{\rm Comp}(\Delta R_c)^3/H_{\rm CAS}m_p$,
which is due to the difference in the gas pressure between the bottom and the top of the clouds, where $m_p$
is the proton mass and $k$ is the Boltzman constant. The gravity
in the vertical direction is given by $F_{\rm G}\sim G\mbh \rho_{\rm c}(\Delta R_c)^3H/R^3$. We find that
\begin{equation}
\frac{F_{\rm gas}}{F_{\rm G}}\sim \frac{\rho_{_{\rm CAS}}}{\rho_{\rm c}}\sim 10^{-3},
\end{equation}
where we use $H\sim c_s/\Omega_{\rm K}$ and $c_s=(kT_{\rm Comp}/m_p)^{1/2}$ and
$\rho_{_{\rm CAS}}/\rho_c\sim 10^{-3}$.

The angular momentum carried by the clouds inevitably causes them to spiral downward towards the disk.
Since cold clouds form in a timescale shorter than the dynamical or the Keplerian timescales, the angular
momentum of cold clouds should follow that of the hot medium. The angular momentum of the hot medium is
given by equation (42). For a given angular momentum value, a cloud's orbit around the SMBH is easily
calculated. Detailed dynamics will be derived in a following paper, from which we will compute emission-line
profiles which can be compared with observations.

\subsubsection{Spatial distribution of the ionization parameter}
Since the density profile varies as $\rho\propto R^{-1.5}$ as shown in Figure \ref{density}, we have
$\Xi\propto R^{\sim -0.5}$. Figure \ref{ionization} shows the ionization parameter in the CAS calculated
from Eq. (13). It shows that the vertical $\Xi-$structure is quite homogeneous whereas the radial structure
can be roughly divided into an inner high-$\Xi$ and an outer low-$\Xi$ region at the boundary
$R_{\rm ion}\sim 10^4R_{\rm Sch}$. At that radius, $\Xi$ falls below a value of about $10^{1.2}$.
This corresponds to $U\sim 0.15$. Photoionization models show that $U>10^{-1}$ for HIL (\civ) regions whereas
$U\sim 10^{-2}$ for LIL (H$\beta$) regions. Photoionization models show
that $U=1-10$ for HIL (\civ) regions whereas $U=10^{-2}-1$ for LIL (H$\beta$) regions (e.g. Marziani
et al. 2010).

As a brief summary of this section, we have shown that there are four phases of the evolving Compton
atmosphere. This drives an
episodic appearance of the broad line regions, which can be compared with observations.
We have systematically analyzed the thermal instability of the Compton atmosphere and find that it is
generally driving the formation
of clouds once the Compton atmosphere has entered Phase II. The key point in the present model is the
continuous injection of warm gas supplied by the star forming disk. The minimum and maximum size of clouds
can be estimated from the instability analysis. The final column density of clouds has been estimated
from the model. Figure \ref{blr} sketches the global scenario described by this model.

\section{Broad line regions}
Since there is a metallicity gradient in the Compton atmosphere (W11), the properties of clouds formed
from that gas also depend on the gradient. Since the thermal equilibrium temperature is higher with higher
metallicity (Fig. 3), the high-metallicity, high-ionization clouds in the innermost regions are hotter
than the low-metallicity, low-ionization clouds found in the outer part of the Compton atmosphere.
The ionization parameter reads $\Xi=L/(4\pi R^2 nkT)\sim N_{\rm H}^{-1}T^{-1}\propto T^{-0.3}Z^{0.5}$,
implying $\Xi$ is a function of distance of the ionized clouds from the center.
We simply divided the Compton atmosphere into High Ionization Line (HIL) regions and Low Ionization Line
(LIL) regions at the break-point $\Xi = 10^{1.2}$, which occurs at $R_{\rm ion}\sim 10^4R_{\rm Sch}$
(Fig. \ref{ionization}).

The separation between the HIL and LIL regions has been well known since the work of
Netzer (1980), Kwan \& Krolik (1979) and Collin-Souffrin et al. (1988). However, these authors only postulated
that the two different BLR regions must exist in order to explain the observed properties of broad emission
lines. Our present model predicts the existence of the two regions, based on physical arguments, as a natural
consequence of the presence of a SF disk. We find that they have spherical and flattened geometry, respectively,
separated at about at $\sim 0.1$pc.

\subsection{Two-phase medium}
As we have shown, continuous injection from the SF disk drives the CAS into phase II, and then the thermal
instability leads to the CAS becoming a two-phase medium consisting of cold clouds and Compton gas. The
filling factor of the evolving BLR can be estimated. The mass of individual clouds is
\begin{equation}
\Delta m_c=\frac{4\pi}{3}\lambda_{\rm TI}^3\ncas m_p=3.5\times 10^{-9}~\lambda_{14}^3n_6~\sunm,
\end{equation}
where $\lambda_{14}=\lambda_{\rm TI}/10^{14}$cm is the initial length of the perturbations. Considering
the fact that the timescale of cloud formation is much shorter than that on which gas is supplied from the
SF disk, when the CAS enters phase II the supplied gas is rapidly converted into clouds. The fraction of
the mass that is converted into clouds can be determined from the ionization parameter $\Xi$. For a CAS
with mass $M_0$, the cloud formation stops if $\Xi\ge \Xi_c$, where $\Xi_c\sim 10$ (Krolik et al. 1981),
and we have the fraction $f_c=M_0/n_{\rm h}Vm_p-1=\Xi_c/\Xi_0-1$, where $\Xi_0=L/4\pi R^2c n_0kT$ and
$n_0=M_0/Vm_p$, namely, mass $M_0$ with a density $n_0$ has the minimum $\Xi$. When $\Xi$ changes by
$\Delta \Xi$, the fraction $f_c=\Delta \Xi/\Xi$, which will be much less than the unity. Since
$t_{\rm sink}\gg t_{\rm cloud}$, we have from Equation (18)
\begin{equation}
\dot{N}_c\approx \frac{f_{0.1}\dot{M}_{\rm inj}}{\Delta m_c}
         \approx 10^8~f_{0.1}\dot{M}_{\rm inj,0}\Delta m_{-9}^{-1}~{\rm yr^{-1}},
\end{equation}
where $f_{0.1}=f_c/0.1$, $\Delta m_{-9}=\Delta m_c/10^{-9}\sunm$, and the accumulated number of clouds
during the Phase II is given by
\begin{equation}
\mathscr{N}_{\rm II}\approx \dot{N}_ct_2=2.6\times 10^{10}~\dot{N}_{c,8}\left(\frac{t_2}{2.6\times 10^2{\rm yr}}\right),
\end{equation}
where $\dot{N}_{c,8}=\dot{N}_c/10^8~{\rm yr^{-1}}$. Such a large number of clouds is enough to explain
the smooth profile of the broad emission lines (Arav et al. 1997). The filling factor can be simply
estimated as
\begin{equation}
{\cal{C}}_{\rm II}\approx \mathscr{N}_{\rm II}\left(\frac{\lambda_c}{R}\right)^2\left(\frac{\lambda_c}{H}\right)
         \approx 10^{-6}~\mathscr{N}_{10}\left(\frac{\lambda_{13}}{R_{\rm 1pc}}\right)^3
                     \left(\frac{H}{R}\right)_{0.1}^{-1},
\end{equation}
where $\mathscr{N}_{10}=\mathscr{N}_{\rm II}/10^{10}$, $\lambda_{13}=\lambda_{\rm cl}/10^{13}$cm
and $(H/R)_{0.1}=(H/R)/0.1$ for the Compton atmosphere.
This filling factor is generally consistent with observations of the line luminosity. This expression
is important for testing the present model through the filling factor of the broad line regions.
It indicates that the ``age" of the BLR in phases II and III can be represented by the filling factor,
which can be estimated from the equivalent widths of the emission lines. We note that such a large
number of clouds cloud lead to collisions among them, which could be described by Boltzmann equation (Whittle
\& Saslaw 1986). We do not treat this problem in this paper.

During Phase II, clouds formed through thermal instability are sinking to the SF disk. The sinking rate
is evolving with time. Without detailed calculations of cloud dynamics, it is not trivial to deduce the
sinking rate. However, the maximum sinking rate can easily be obtained from
\begin{equation}
\dot{N}_{\rm sink}\lesssim \dot{N}_c \approx 10^8~ {\rm yr^{-1}}.
\end{equation}
This rate can be tested from observations of AGNs with redshifted intermediate components of H$\beta$
(Hu et al. 2008b).

The behavior of the clumpy BLR in Phase III can be estimated in a way similar to the above. The BLR is
full of cloudlets emitting broad emission lines, but the intermediate component of H$\beta$ shifts backward
toward the rest wavelength until it overlaps with the very broad component, with an observable absence of
the redshifted component. We point out that different parts of the BLR may be in different phases at the
same time due to the variation of the cloud and hot-phase gas properties as a function of distance from
the black hole.

\subsection{The high-ionization-line BLR: outflowing clouds}
In the presence of the metallicity gradient, the innermost region of the SF disk usually has metallicity
$Z \sim 10Z_\odot$ (W11). For a concise discussion, we take the metallicity $Z=10Z_\odot$ in this region.
The thermal equilibrium temperature is $T \lesssim 6\times 10^4$K (from Figure \ref{HC} {\em left} panel).
The column density of the clouds formed here is about $N^{\rm H}_{\rm HIL}\sim 5\times 10^{21}Z_1^{-0.5}$
from equation (63). This simply indicates that the clouds in the HIL region are optically thin. The
gravitational acceleration is given by $\gbh=G\mbh/R^2=15.0~M_8^{-1}r_3^{-2}{\rm cm~s^{-2}}$.

The cold clouds are accelerated by radiation pressure. For optically thin clouds the strength of the
radiation pressure depends linearly on the metallicity as $a_{\rm rad}\propto Z$, but for partially-ionized
clouds the dependence is approximately $a_{\rm rad}\propto Z^{0.4}$ (Abbott 1982), where $a_{\rm rad}$ is
the acceleration of radiation due to line absorption. We simply take a linear dependence for optically
thin clouds and neglect radiative acceleration for optically thick clouds. The metallicity gradient in
the BLR leads to difference of a factor of a few in the radiation force multiplier for the LIL and HIL
regions. The acceleration due to radiation pressure of an isotropic field is given by (e.g. Netzer \&
Marziani 2010)
\begin{equation}
a_{\rm rad}(R)=\frac{\sigma_{\rm T}L_{\rm Bol}}{4\pi R^2cm_p}{\cal M}
           =108.0~\ell_{0.3}\alpha_{0.2}M_8^{-1}r_3^{-2}Z_0N_{22}^{-1}~{\rm cm~s^{-2}},
\end{equation}
where $N_{22}=N_{\rm H}/10^{22}{\rm cm^{-2}}$,
${\cal M}=(Z/Z_\odot)\alpha_{\rm ion}/\sigma_{\rm T} N_{\rm H}$ is the force multiplier, and
$\alpha_{\rm ion}$ is the fraction of the bolometric luminosity absorbed by the clouds. Here we
use $\alpha_{0.2}=\alpha_{\rm ion}/0.2$, which is absorbed by the gas.
We note that this acceleration is comparable with the gravity of SMBHs, but is dominant for
$N_{\rm H}<10^{22}{\rm cm^{-2}}$. The accelerated cloud will escape the potential of the SMBH
if it reaches the escape velocity $V_{\rm esc}=(G\mbh/R)^{1/2}=6.7\times 10^3~r_3^{-1/2}\kms$
for which the timescale is
\begin{equation}
t_{\rm acc}=\frac{V_{\rm esc}}{a_{\rm rad}}=0.2~M_8r_3^{3/2}\ell_{0.3}^{-1}
            \alpha_{0.2}^{-1}Z_0^{-1}N_{22}~{\rm yr}.
\end{equation}
The timescale for then actually escaping is given by $t_{\rm esc}=R/V_{\rm esc}=1.4~r_3^{3/2}M_8~{\rm yr}$,
which is comparable with the acceleration timescale. The escape timescale is much shorter than the Keplerian
rotation period. Therefore, optically thin clouds clouds are very rapidly blown away by the radiation pressure
as soon as they are born.

The HIL-BLR is dominated by optically thin clouds, and hence by the outflowing clouds. The observed \civ\ and
other HILs often show complicated profiles, indicating outflows. The mass rates
can be estimated from the supply rates from the star-forming disk within the HIL regions.
The clouds could be transported to the narrow line regions with a timescale of
$t_{\rm tran}=R/V_{\rm esc}=1.6\times 10^5~R_{\rm 1kpc}V_{6000}^{-1}$yr, where $R_{\rm 1kpc}=R/1{\rm kpc}$
and $V_{6000}=V_{\rm esc}/6000\kms$. During the transportation,
the clouds will appear as warm absorbers in soft X-rays, which are observed in many objects.

{
\centering
\figurenum{8}
\begin{figure*}
\centerline{\includegraphics[angle=-90,scale=1.0]{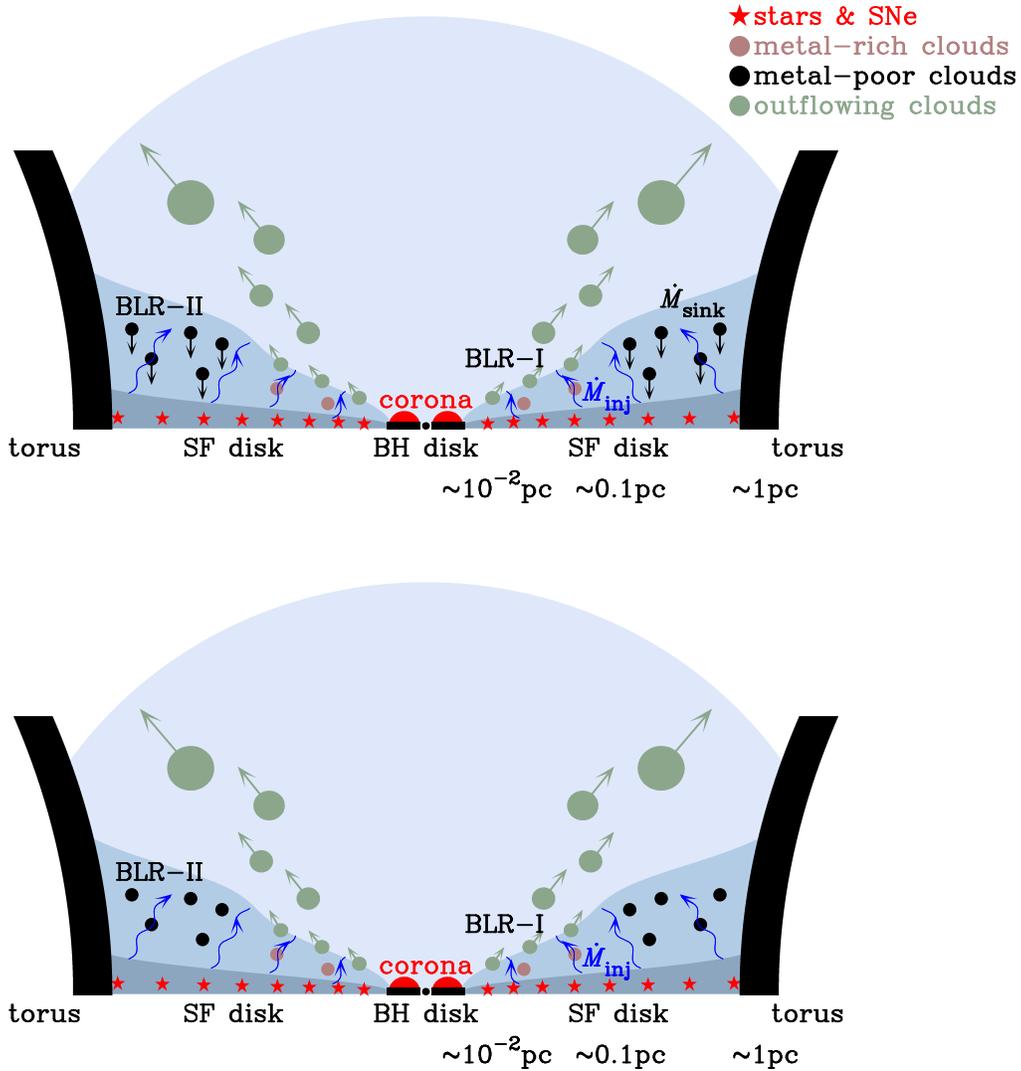}}
\figcaption{\footnotesize Sketches of the BLR in phase II ({\em upper panel}) and phase III ({\em lower panel}).
In either phase, the BLR extends inwards from the inner edge of the dusty torus to the self-gravity radius
of the accretion disk. SNexp heating causes gas to be injected from the SF disk at a rate $\dot{M}_{\rm inj}$.
This gas is heated by emission of the SMBH accretion disk to become the Compton gas, and then forms into discrete
clouds due to the two-phase instability in the Compton atmosphere. BLR-I is formed from the metal-rich gas that
is injected from the inner part of the SF disk and then is pushed outwards and upwards away from the disk by
radiation pressure. BLR-II is formed from the lower-metallicity gas injected from the outer part of the SF disk,
which then stays close to the disk surface. In the case of phase II ({\em upper panel}), individual clouds form
in BLR-II and then sink back down into the SF disk. In phase III ({\em lower panel}), these clouds are dynamically
destroyed and merge back into the Compton atmosphere before they have time to sink all the way to the SF disk.
}
\label{blr}
\end{figure*}
}

\subsection{Low ionization line BLR: transient states}
For the LIL-BLR, the ionization states of clouds are relatively lower since they are located further
away from the central energy source of the SMBH accretion disk. On the other hand, the metallicity of
clouds originating in the outer part of the SF disk is lower than the innermost regions. W11 show
that the metallicity is about $0.1Z_\odot$ in this region. When the CAS enters phase II, the new
thermal equilibrium of clouds with low metallicity results in a temperature around $10^4$K (from
Figure \ref{HC} {\em right} panel). The final column density of the clouds is then
$N_{\rm LIL}^{\rm H}\sim 5.4\times 10^{22}Z_{0.1}^{-0.5}~{\rm cm^{-2}}$. Such a thick cloud only
undergoes negligible radiative acceleration.
The dynamics of these clouds is relatively simple in phase II. They are spiraling downward towards
the SF disk as shown by equation (65). The timescale of sinking to the SF disk is
$t_{\rm sink}\sim t_{\rm Kep}=R/V_{\rm K}$. These sinking clouds constitute
infalling flows, showing redshifted Balmer lines to the observer.
Later, after further continuous injection from the SF disk, the Compton atmosphere becomes so
dense that it prevents the clouds from sinking to the SF disk. The CAS then enters phase III. This
results in much smaller infall velocities, and the observed Balmer lines shift back toward their rest
wavelength. Finally the LIL regions collapse into the SF disk since the entire CAS becomes dominated
by line cooling. These transitions are qualitatively discussed in \S3.1.

We would like to stress the difference between the mass circulation in phase II and the mass cycle in
phase III. The former term refers to the circulation between the CAS and SF disk (i.e. the sinking of
clouds back onto the SF disk), leading to a slower growth of the CAS than in phase I.
The term ``mass cycle", on the other hand, refers to a rapid exchange of mass back and forth between
the CAS and the clouds due to the rapid dynamical destruction of clouds as they move through the CAS.
This latter cycle drives the CAS to rapidly become sufficiently dense that it enters phase IV, the
collapse of the BLRs.

In summary, we have attempted to build up a global scenario of BLRs working from first principles. The
key ingredient in this model is the continuous injection of warm gas from the SF disk. This gives rise
to the episodic appearance, disappearance, and then the reappearance of the broad line regions, with four
phases within each episode. The fates of clouds formed in different regions are determined by the
metallicity of the gas coming from different parts of the SF disk. Low and high ionization regions
are then formed naturally. The HIL clouds are blown away by the radiation pressure, transporting
metal-rich material to the narrow line regions. The LIL region undergoes a more complicated evolution,
finally resulting in the collapse of the entire Compton atmosphere. In light of the complexity of the
above analytical discussions, it would be worth carrying out numerical simulations
to show more details of these phases of BLR formation.

\begin{deluxetable*}{lllll}
\tabletypesize{\scriptsize}
\tablewidth{0pt}
\tablecaption{{\sc Predicted Properties of Evolving BLRs in a Single Episode}}
\tablehead{
Phase & Observed properties      &   Duration        & Appearance & Notes \& Ref.\\
      &                          &   (Years)         &            &          }
\startdata
I     & bright type II-like AGNs without polarized broad lines    & 25.3             & $\sim  1.6\%$ & a few objects\\
II    & one VBC and one redshifted-IMC and redshifted Fe {\sc ii} & 260.0            & $\sim 16.0\%$ & H08\\
III   & overlapped one VBC and one IMC, non-redshifted Fe {\sc ii}& $1.3\times 10^3$ & $\sim 79.7\%$ & H08\\
IV    & appearance of the maximum equivalent width of lines & 45.1    & $\sim  1.8\% $ & ???\\
\enddata
\tablecomments{VBC: very broad components with a typical width of a few $10^3\kms$, and IMC: intermediate
component with a typical width of $\sim 10^3\kms$. Evolution of AGN BLR from phase I-IV forms
a spectral sequence of BLR as indicated by the observed properties. ``A few objects" refers to
NGC 3660 (Tran 2001); 1ES 1927-654 (Boller et al. 2003); ESO 416-G00211 and PMN J0623-4636
(Gallo et al. 2006); Q2130-431 (Panessa et al. 2009) and two SDSS quasars: J161259.83+421940.3 and
J104014.43+474554.8. H08: Hu et al. (2008a,b).}
\label{summary_prediction}
\end{deluxetable*}

\section{Observational tests of the episodic BLR}
The present model makes a number of clear theoretical predictions: 1) the BLRs are transient; 2) the
episodes of BLR state transition, driven by the star formation in the self-gravitating disk, are a few
thousand years in length, giving rise to a spectral sequence of broad emission lines; 3) the BLRs are
separated into HIL and LIL regions, which form a steady gradient of metallicity, with HIL regions
having higher metallicity than LIL regions; and 4) there is an intrinsic
connection between the BLR and NLR through outflows developed from the HIL regions.

These predictions can be tested observationally. First of all, non-BLR AGNs are a key constituent
of the transient BLRs. Here non-BLR AGNs are those that have typical accretion rates ($\sim 0.2L_{\rm Edd}$),
but do not have BLRs. These represent phase I in our model. We have to find them, even though they are
relatively rare (only $\sim 10^{-2}$ probability of appearance; see Table 2). Second, we should be able
to find and identify QSOs in each of the sequential phases predicted by our model -- there should be
a spectral sequence of broad emission lines. Third, although the existence of separate HIL and LIL BLRs
has been known observationally for many years, we predict that the HIL and LIL regions have metallicity
differences arising from the different rates of star formation in different parts of
the SF disk. The observed metallicity difference between HIL and LIL regions (Warner et al. 2003; 2004)
lends support to the prediction of our model. Detailed theoretical discussions can be found in W11. Fourth,
the present model clearly predicts an intrinsic connection between BLR and NLR through outflows. This
section is devoted to discussing these key tests from observations.

\subsection{Observational appearance during phase I}
The diversity of AGNs is often attributed mainly to differences in the orientation of a dusty torus with
respect to the observer's sight line (Antonucci 1993). Considerable evidence has been found in favor of
this scenario, such as the presence of polarized BLRs in some Seyfert 2 galaxies and the larger amount
of absorbing material in Seyfert 2s observed at X-ray wavelengths. However, over the last decade there
has been increasing evidence suggesting that this orientation-based unification
scheme is not the whole story. Some optically-identified type I AGNs have quite large absorption
in the X-rays (Fiore et al. 2001; Mateos et al. 2005; Cappi et al. 2006).
Some type II AGNs are found without X-ray absorption (Pappa et al. 2001;
Panessa \& Bassani 2002; Barcons et al. 2003; Caccianiga et al. 2004; Corral et al. 2005; Wolter et al.
2005; Bianchi et al. 2008; Brightman \& Nandra 2008). These results directly confront the simple
version of the dusty torus model, and have motivated suggestions of evolving
broad line regions and/or varying accretion rates in addition to the orientation-based effects (Panessa
\& Bassani 2002; Wang \& Zhang 2007). On the other hand, the complex range of observed properties of
intermediate Seyfert galaxies, such as Seyfert 1.8 and Seyfert 1.9 (Trippe et al. 2010), implies that
non-Seyfert 1 galaxies are composed of intrinsically different populations. It has been realized
that 1) partial obscuration by a clumpy torus may explain the transition from type II to I; 2) low
accretion rates can dilute the clouds in the broad line regions; and 3) abnormal gas-to-dust ratios
in the torus can cause AGN to appear as type II objects at optical wavelengths, but without significant
absorption of X-rays. It is plausible that the phase I objects predicted by the present model appear among
these abnormal objects.

For convenience, Table \ref{unification} lists the known types of AGNs and the likely parameters of
their central engines. In the following subsections, we discuss the relationship of these different
types of AGN to the predictions of the present model.

\begin{deluxetable*}{lllll}
\tabletypesize{\footnotesize}
\tablewidth{0pt}
\tablecaption{{\sc AGN Types in the Unification Scheme}}
\tablehead{
Type  & Eddington ratio  & SMBH mass ($\sunm$)& $N_{\rm H}({\rm cm^{-2}})$ & Note \& Ref.\\
~~(1)        &  ~~~~~~(2) &  ~~~~ (3)        &~~~~ (4)      & ~~~(5)}
\startdata
Seyfert 1            & $\sim 0.25$ & $\sim 10^7$  & $\lesssim 10^{22}$& $\surd$\\
Seyfert 2            & $\sim 0.25$ & $\sim 10^7$  & $\gtrsim 10^{22} $& $\surd$\\
``True" Seyfert 2    &$<10^{-2}$   & $\sim 10^7$  & $<10^{21}$        & $\surd$\\
Narrow line Seyfert 1&$\sim 1$     & $10^6-10^7$  & $<10^{22}$        & $\surd$\\
Obscured NLS1s       & $\sim 1$    & $10^6-10^7$  & $\gtrsim 10^{22}$ & $\surd$ (ZW06)\\
Unabsorbed Seyfert 2 & $\sim 0.25$ & $\sim 10^7$  & $\lesssim 10^{22}$& $\surd$ \\ \hline
\multicolumn{5}{c}{New types predicted by the present model}\\ \hline
Phase-I AGNs         & $\sim 0.25$ & $\sim 10^7$  & $\lesssim 10^{21}$& ``Panda" AGNs, ? \\
Phase-IV AGNs        & $\sim 0.25$ & $\sim 10^7$  & $\lesssim 10^{21}$& very large EW, ?
\enddata
\tablecomments{
Columns are: (1) types of AGNs (see details in Wang \& Zhang 2007); (2) the averaged
Eddington ratio;
(3) SMBH mass in units of solar mass; (4) absorbing hydrogen column density; and (5) notes.
Nomenclature: NLS1 = narrow line Seyfert 1; ZW06 = Zhang \& Wang (2006); ``$\surd$" indicates that this type AGN
has been found; ``Panda" AGNs are those Seyfert 2 galaxies that have no BLR, but have
high accretion rates ($\sim 0.25$); ``?" means that their existence is unconfirmed, and that we
suggest searching for them in the SDSS. The blue types of AGNs are predicted by the present paper.}
\label{unification}
\end{deluxetable*}

\subsubsection{Panda AGNs: non-BLR AGNs with normal accretion rates}
It is often assumed that all type II AGNs have hidden BLRs. This idea is based on observations of
Seyfert 2 galaxies that in polarized light show broad Balmer lines due to scattered light coming
from a BLR obscured by a dusty torus (Antonucci \& Miller 1985; Antonucci 1993). Recently, it has been
discovered that some type II AGNs show a deficit of the polarized broad lines (Gu \& Huang 2002; Nicastro
et al. 2003), and have accretion rates lower than a critical value of $L_{\rm bol}/L_{\rm Edd}\sim 10^{-3}$
(Nicastro et al. 2003; Wang \& Zhang 2007; Elitzure \& Ho 2009). This has  been confirmed by further
observations (Tran et al. 2011). These objects are called ``true" type II AGNs, or unabsorbed non-hidden
BLR AGNs. The present model predicts a new type of AGNs, which have normal accretion rates and no BLRs.
Since phase I is relatively short, AGNs in this state are rare. We designate those as ``Panda"
AGNs\footnote{Our term ``Panda" is meant to convey the idea these are a rare and hard-to-find species,
but like their namesakes in the animal kingdom, one that nevertheless does exist} in order to distinguish
them from objects with low accretion rates.
There is no doubt that the existence of this kind of AGN is a key test of the present model.

Wang \& Zhang (2007) systematically examined a broad sample of published AGN spectra in the light of the
unification scheme and the evolutionary influence of SMBH growth. One object of particular interest in
their sample is  NGC 3660, which shows no polarized broad lines (Tran 2001). Its \oiii\ luminosity is
estimated to $6.5\times 10^{41}$\ergs (Kollatschny et al. 1983). With the bolometric luminosity of
$L_{\rm Bol}\sim 10^{45}$\ergs, along with the SMBH mass $\mbh\sim 10^7\sunm$ estimated from the Magorrian
relation (Magorrian et al. 1998), we find an Eddington ratio of about unity. NGC 3660 is the
only such object in the Wang \& Zhang (2007) sample.
Panessa et al. (2009) have more recently discussed an interesting Seyfert 1.8 galaxy, Q2130-431, that has
weak broad H$\alpha$ and H$\beta$ emission. It has an Eddington ratio of $L_{\rm Bol}/L_{\rm Edd}\sim 0.4$,
absorption column density $N_{\rm H}<10^{20}{\rm cm^{-2}}$ and a Balmer decrement of 0.34 in light of
H$\beta$/H$\alpha$ ratio, indicating that
the BLR is not suffering from heavy reddening and is not obscured. These authors concluded that the weakness
of BLR in Q2130-431 is {\em intrinsic} in spite of the high accretion rates. This object seems to challenge
the popular scenario that the absence of a BLR is driven by the low Eddington ratios. However, this could
be clear evidence for the presence of a transient BLR in phase I suggested by the present model. Though the
completeness of the Wang \& Zhang (2007) sample combined with
the results of Pennesa et al. (2009) is uncertain, the ``Panda"
AGNs constitute only $\sim 2/245= 0.8\%$ of the population, which is consistent with the prediction of the
present model for the fraction of AGNs in phase I (see Table \ref{summary_prediction}).

There is increasing evidence for the presence of type II AGNs with low X-ray absorption and high accretion
rate, for example, 1ES 1927+654 with $N_{\rm H}\sim 7\times 10^{20}{\rm cm^{-2}}$ and
$L_{\rm 0.1-2.4keV}\sim 5\times 10^{43}$\ergs (Boller et al. 2003), and ESO 416$-$G00211 type II AGNs with
$N_{\rm H}<3\times 10^{20}{\rm cm^{-2}}$ and $L_{\rm 2-10keV}\sim 4\times 10^{43}$ and PMN J0623-4636
with $N_{\rm H}\sim 3\times 10^{19}{\rm cm^{-2}}$ and $L_{\rm 2-10keV}\sim 2\times 10^{44}$\ergs
(Gallo et al. 2006). These typical Seyfert galaxies contain SMBHs with $10^7\sunm$, they thus have
Eddington ratios between $0.1$ and 1 as determined from their X-ray luminosities. Though the
reason why type II AGNs have such low absorption remains a matter of debate, they are candidates
to be the ``Panda" AGNs in the present model. Future polarization
observations of these objects should be made to search for obscured BLRs.

\subsubsection{Type II quasars}
Type II quasars are the more luminous analogues of the type II AGN discussed above, showing only narrow
emission lines. They are rare compared to type I quasars, but a considerable number of them have been
found in the large SDSS QSO sample(e.g. Zakamska et al. 2003; 2005). Do these include Panda-type quasars?
Colors of most of these objects fall between those of galaxies and quasars (Zakamska et al. 2003), implying
a nuclear continuum obscured by a dusty torus together with underlying continuum contamination by their host
galaxies. {\em Chandra} observations of 12 type II quasars find that they generally have high absorbing column
density, $N_{\rm H} > 10^{22}{\rm cm^{-2}}$ (Vignali et al. 2006), but this is a very limited sample. According
to the model, non-BLR ``Panda" quasars should be detected with a probability of $10^{-2}$ from the SDSS.

\begin{figure*}[t!]
\centering
\figurenum{9}
\includegraphics[angle=0,scale=0.85]{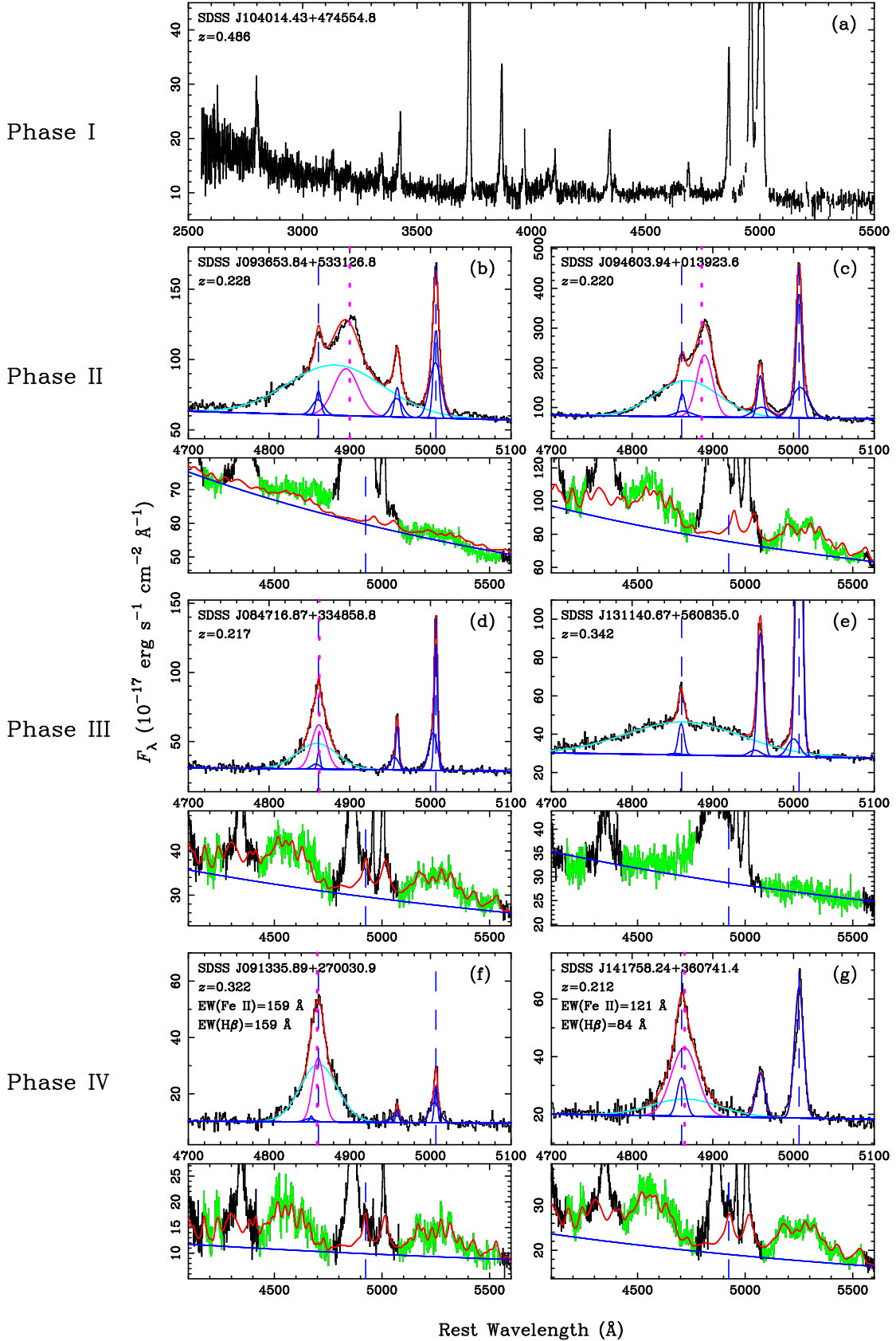}
\figcaption{\footnotesize (a) A candidate phase I AGN, from SDSS. The Eddington ratio is about
$L_{\rm Bol}/L_{\rm Edd}\sim 0.9$. See text for a detailed explanation. $(b - g)$ Spectral sequences
of the broad emission line profies corresponding to phases II, III and IV as indicated.
A redshifted component of H$\beta$ appears in phase II and disappears in phase III. Emission lines
reach their maximum equivalent width and the BLR is ready to collapse in phase IV. The appearance
percentage of each phases can be found in Table 2, agreeing with the observational statistics from Hu
et al. (2008a,b).}
\label{spectral_sequence}
\end{figure*}

We made a further search for these Panda-type objects in the SDSS sample from SDSS DR7 (those objects are actually
type II quasars in light of Zakamsa et al. 2003; 2005), using the criteria of color and line width. We
give an example of those objects labeled by Phase I as shown in Figure \ref{spectral_sequence}. This
quasar has a weak \oii\ line and hence the continuum is less contaminated by young stars.
The \oiii(FWHM) is about $660\kms$ and the dispersion velocity $\sigma$=FWHM(\oiii)/2.35$\sim 280\kms$. We have
$\mbh\sim 10^{8.7}\sunm$ from $\mbh-\sigma$ relation (Tremanine et al. 2002). From the \oiii\, luminosity of about
$10^{43}$\ergs, we find its bolometric luminosity to be
$L_{\rm Bol}\sim 3500L_{\rm [OIII]}\approx 6\times 10^{46}$\ergs and the Eddington ratio
to be $L_{\rm Bol}/L_{\rm Edd}\sim 0.9$.  This object has a normal quasar luminosity as
measured from its continuum, implying that its central engine is viewed directly, but that there is
no BLR. Our model can explain the absence of the BLR in quasars and AGNs with normal Eddington ratios.

We have to point out that identification of ``Panda" quasars should deal with other properties as argued
for ``Panda" AGNs in \S5.1.1. In principle, some of them should resemble to weak emission line quasars
(Shemmer et al 2010). In the future it will be well worth carrying out a systematic search for
the Panda-type quasars predicted to exist in the SDSS, and then measuring the properties of those AGNs.

\begin{figure*}[t!]
\centering
\figurenum{10}
\includegraphics[angle=-90,scale=0.65]{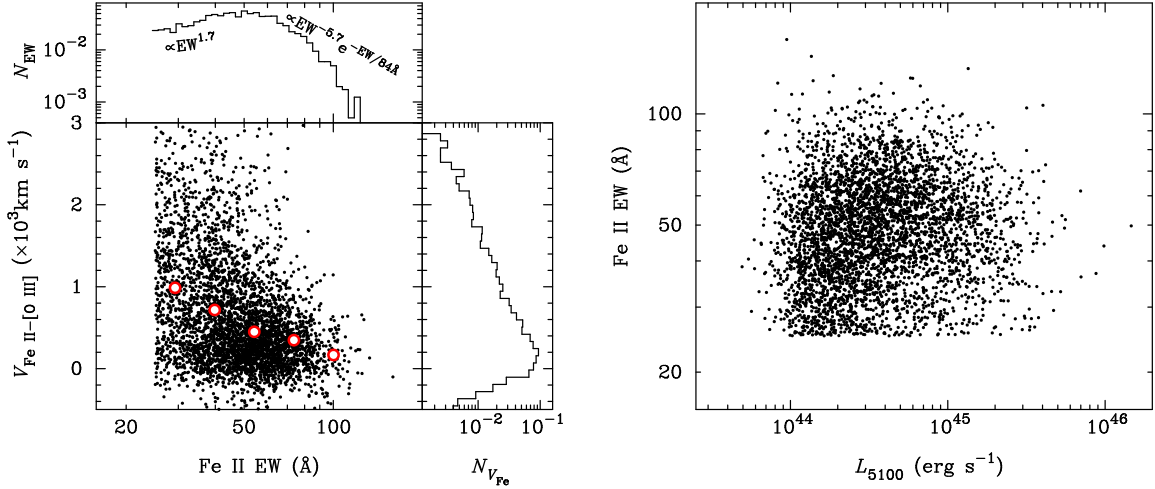}
\figcaption{\footnotesize
{\em Left panels}: The sinking velocity of Fe {\sc ii} clouds versus the equivalent width. The \feii\
equivalent width is obtained by EW=$F$(\feii)/$F_{5100}$, where $F$(\feii) and $F_{5100}$ are
fluxes of \feii\ between 4434\AA\ and 4684\AA, and 5100\AA, respectively. The density of
Compton gas is proportional to the number density of clouds, implying the EW as an indicator of covering
factor. We use the EW as an indicator of the CAS density. The data show an anti-correlation of the velocity
with the EW, which is consistent with the model. We divided the EW of \feii lines into five bins. The red
circles around white interiors show the averaged value of the infalling velocity of clouds in each bin.
{\em Upper-left panel}: the distribution of the EW(\feii) normalized by the total of 4037 objects. The
very steep distribution of high EW implies the presence of objects with maximum EW. The
$N_{V_{\rm Fe}}-$panel shows the distribution of \feii\, cloud velocity. {\em Right panel}: Test to
determine if there is an anti-correlation between EW(\feii) and the ionizing luminosity $L_{5100}$. The
absence of a correlation implies that the maximum EW of \feii~lines is an intrinsic feature rather than
being caused by the flux limit of SDSS survey.}
\label{sink}
\end{figure*}

\subsection{Appearance of evolving BLRs: a spectral sequence}
The observed broad emission lines from AGNs and quasars exhibit a wide range of central line shifts,
profiles and equivalent widths (e.g. Hu et al. 2008a; 2008b; Marziani et al. 2010). Here we attempt to build
up a spectral sequence of broad lines in light of the BLR states predicted for phase II-III-IV in our
model.

Based on the H$\beta$ profiles, Hu et al. (2008b) roughly divided a sample of approximately 50000 SDSS
quasars into three classes: 1) objects with redshifted intermediate and broad H$\beta$ components; 2)
ones with non-shifted intermediate and broad H$\beta$; 3) ones with only broad components of H$\beta$.
Finally the maximum equivalent width of H$\beta$ in this sample is $EW \sim 500$\AA. Figure \ref{spectral_sequence}
shows the spectral classes of SDSS quasars. Figure \ref{spectral_sequence}{\em a} shows an example of a
possible phase I AGN, as discussed in the previous section. Figures \ref{spectral_sequence}{\em b} and
\ref{spectral_sequence}{\em c} show the H$\beta$ lines with redshifted intermediate components which
characterize class 1 objects. We interpret these as corresponding to phase II of our model, during
which cold clouds return to the SF disk due to the mass circulation between the CAS and the disk. From
panel {\em b} to panel {\em c}, the redshift of the intermediate H$\beta$ decreases. This corresponds to
an increase in the CAS density, leading to a decrease in the infall velocity. When the infall velocity
tends to zero, the BLR enters phase III. We associate this phase with the Hu et al. (2008b) class 3 objects,
such as those shown in Figures \ref{spectral_sequence}{\em d} and \ref{spectral_sequence}{\em e}.
Due to the strong dynamical destruction process, cold clouds are not able to survive long enough to sink
into the SF disk. Finally, in Figures \ref{spectral_sequence}{\em f} and \ref{spectral_sequence}{\em g},
we show the two objects from H08 with the maximum H$\beta$ equivalent width. These two quasars are believed
to be in phase IV, i.e. the collapsing state of the BLR evolution.

We can use the relative numbers of objects to test these proposed connections between the classes observed
by Hu et al. and the phases predicted by our model. The sample of Hu et al. (2008a, b) shows that the fraction
of class 1 objects is $11\%$, while $89\%$ of the sample are class 2) and 3) objects. Based on the timescales
predicted by our model, we predict that phase II objects should comprise roughly $t_2/(t_1+t_2+t_3)\sim 16\%$
of all AGNs, while phase III objects should make up about $t_3/(t_1+t_2+t_3)\sim 80\%$ (Hu et al. 2008b).
This is quite good agreement between the model and the observations, supporting an interpretation that the
empirical classes found by Hu et al. really do represent the progression through the evolutionary phases
predicted by our model.

We note that the \feii\, emission lines usually have properties similar to those of the intermediate
H$\beta$ components in the sense that the two have consistent redshifts and FWHM, suggesting they
originate from the same regions. However, there is still a fraction of objects which show \feii\,
profiles that are different from those of the intermediate H$\beta$ component. Some objects have blueshifted
intermediate H$\beta$, while \feii\, shows redshifts. We suggest that this is due to the complicated
effects of the optical depths in these different lines (Ferland et al. 2009). The optical-passband
\feii\, lines may be better indicators of the LIL regions than H$\beta$ since the former are
emitted isotropically while the later is strongly inward beamed at higher optical depths.

In order to further test a transition from phase II to III, we make use of SDSS data to show the dependence
of infalling velocity on the CAS density.  Our model predicts a steady increase in the density of the CAS until
the last phase of collapse, namely dynamical friction of clouds with the Compton gas increases, and
the covering factor is increasing with time monotonically. This causes the maximum sinking velocity of the clouds to
decrease as the covering factor increases. Since the \feii\ photons from the emitting clouds are more isotropic
than H$\beta$ (Ferland et al. 2009), the equivalent width of the Fe II lines could be a good
indicator of the BLR covering factor since the reprocessed emission will be isotropic for Fe II lines (while for
H$\beta$ formed at large column densities it will be anisotropic). We predict an inverse
correlation between the EW of Fe {\sc ii} and its redshift (the sinking velocity to the SF disk).
Figure \ref{sink} tests this prediction, using data from our systematic study of the properties of H$\beta$
and Fe {\sc ii} lines in SDSS QSOs (Hu et al. 2008a,b). The available
data show large scatter, but the averaged points show a clear trend. The sinking velocity does anti-correlate
with the covering factor, as predicted.

Finally, we consider the last stage of BLR evolution. The large EW of H$\beta$ and \feii\, are the
unique observable characteristic of phase IV. For equivalent widths exceeding some critical limit,
the entire CAS will be cooled through line cooling. From the \feii\, lines, we have the covering
factor of the \feii\, clouds as ${\cal C}\sim 100\%$. We identify them in the last stage of the BLR.
Furthermore, the upper-left panel in Figure 10 shows the observed distribution of the Fe II equivalent
widths. We find that it follows $N_{\rm EW}\propto {\rm EW}^{1.7}$ for sources with low EW whereas
for high EW it has the dependence $N_{\rm EW}\propto {\rm EW}^{-5.7}\exp\left(-{\rm EW/84\AA}\right)$.
We have a cutoff EW of ${\rm EW_{\rm cutoff}}=84$\AA, indicating that there is a maximum EW of
emission lines, corresponding to phase IV. Sources with ${\rm EW >EW_{cutoff}}\sim 84$\AA\, will be
in phase IV. They constitute about $4\%$ of the total objects in the Hu et al. sample. This fraction
agrees well with the ($\sim 1.8\%$) predicted by the present model (Table 2).

We have tested to see if the maximum EW could be a selection effect due to the flux limit of the SDSS
survey. In Figure 10 (right) we plot the potential correlation of EW(\feii) with the ionizing luminosity.
The non-correlation between the two directly shows that the maximum is not caused by the flux-limit effects.
We would like to stress that the clouds are optically thin for \feii\ photons, making \feii\ EW a robust
indicator of the covering factor. The maximum \feii\ EW indicates the last stage (phase IV) of the transient
BLR.

\subsection{Intrinsic connection between NLR and BLR: outflowing clouds}
The cloud outflows which we predict will be ejected from BLR-I by radiation pressure have the important
implications that the clouds may undergo expansion and become highly ionized gas (HIG), which produces many
features in X-rays. Although absorption features due to HIG were first clearly seen in AGN with {\em ROSAT}
(Turner et al. 1993), most of the fundamental physical properties of this component remain a matter for
debate. The HIG appears as absorption lines, such as \ovii\, (0.74keV) and \oviii\, (0.87keV) in type I AGNs
(Reynolds 1997, George et
al. 1998, Kaspi et al. 2000) whereas it is seen as emission lines in type 2 AGNs (Turner et al. 1997, Netzer
et al. 1998). The HIG is variously suggested to be intrinsically related with BLR clouds (Reynolds \& Fabian
1995), or to have evaporated off ``bloated stars" (Netzer 1996), or to be gas evaporated off the dusty torus
obscuring the nucleus (Krolik \& Kriss 1995), or to be a wind driven off the accretion disk (Bottorff et al.
2000). In at least one object, MR 2251-178, the general characteristics of the UV and X-ray absorbers are
consistent with them having originated in the same gas (Kaspi et al. 2004). A brief summary is given by
Risaliti (2010). The dynamics of the radiatively accelerated HIL clouds predicted by our model is worth
investigating to see if these clouds really can be the progenitors of the observed HIG, and it will be
very important to determine the metallicity of the HIG through X-ray spectroscopy. Future work on the
comparison of warm absorbers with the properties of clouds in the HIL BLR is also well worth carrying out
to examine the prediction of the present model through the observational tests of ultraviolet \civ\,
absorption lines with X-ray absorption.

Furthermore, we find a strong correlation between the BLR and NLR metallicity in a sample of quasars
observed with {\em HST} (Wang et al. 2011 in preparation). This lends strong support to there being a
connection between these two regions, even though they are separated by the large spatial distance of
$\sim 1.0$kpc.

\subsection{Other observational tests}
Another firm prediction of our model is that the HIL lines (BLR-I) will have higher metallicity than the LIL
lines (BLR-II). Metallicity in quasars can be estimated by measuring the N/C and N/O abundance ratio by using
the \nvciv, and \niiiciii\ methods described by Hamann \& Ferland (1999), Hamann et al. (2002) and  Baldwin
et al. (2003b). As we have
discussed in Paper I, the expected differences may have already been found, with $Z_{\rm HIL}\sim 5Z_{\rm LIL}$
from the stacked spectra (Warner et al. 2004). Additional checks of these results are badly needed.

Finally, future reverberation mapping of the AGNs that are thought to represent different phases in our model
could provide observational tests of the model(like Denney et al. 2009; 2010). The self-gravitating part
of the accretion disk has been
regarded as a likely source of the BLR for a quite long time (e.g. Collin \& Hure 2001; Bian \& Zhao 2002).
The argument is favored by the fact that the self-gravitation radius is about the same as the reverberation
mapping radius. Our results are also generally compatible with the well-known result from
reverberation mapping that the HIL lines are formed at smaller radii and the LIL lines at larger radii
(Peterson \& Wandel 1999), but a detailed comparison is beyond the scope of this paper. Mont-Carlo simulations
of the reverberation mapping relation will be carried out in a separate paper.

\begin{sidewaystable*}
\vglue -9cm
\caption{Summary of BLR geometry and dynamics in AGNs}
\centering
\begin{tabular}{l|llll}
\hline\hline
Classes & References & Dynamics & Geometry & Notes \\
\hline
General & W59    & giant potential of gravity & sphere     & seminal paper of the broad H$\beta$ line \\
        & KMT81  & cloud-intercloud system in thermal balance & HIM & \\
        & C88    & & disk (LIL) + sphere (HIL) &  \\ \hline
Model A & S85    & clouds developed from RDR region of AD& disk-like & 3 regions of SS disk\\
        & BMS83  & Compton heated winds from AD & ... & ... \\
        & SR85   & Compton heated winds from AD & ... & ... \\
        & RNF89  & Clouds bound by magnetic fields & ... & ... \\
        & W96    & Compton heated winds from AD & ... & ... \\
        & CH01   & outer part of accretion disk & ... & ... \\ \hline
Model B & H75    & tidal disruption of stars & outflow &  \\
        &S77     & disk winds driven by emission of inner disk & disk & outflow mass rate $10^{-6}\sunmyr$ \\
        &BM75    & outflowing clouds driven by AD radiation & sphere & Logarithmic profile \\
        &C80     & ...& outflow + rotating disk &  \\
        &E80     & 'Cometary Star' tail & cusp-like & \\
        &N80     & & HIL \& LIL & stratified layer with  \\
        &B81     & outflows & spherical geometry & \\
        &WMY85   & outflows & quasi-spherical geometry & \\
        &SN88    & SMBH potential & winds from stars & mass loss of red giant stars with $10^{-5} \sunmyr$\\
        &R92     & tidal disruption of stars & unbound+bound clouds & hot + cold clouds \\
        &T92     & evolved SN remnant  & ...           & \\
        &CR96    & surface of accretion disk & ... & \\
        &MC97    & AD-driven outflows & &  \\
        &N03     &$p_{\rm gas}\sim p_{\rm rad}$ transition regions& &$p_{\rm gas}$ and $p_{\rm rad}$ are gas and radiation pressures, respectively.\\
        &CH11    & radiation pressure-driven winds & & is partially favored by blue-shifted H$\beta$\\ \hline
Model C &EBS92   & magnetized winds from disk & & \\ \hline
Model D &ALW93   & stars interact with accretion disk & & \\
        &VC02    & star-disk \& star-star interaction & inflow & \\
        &ZSC94   & star-disk interaction & star tail above accretion disk &  \\ \hline
Model E &AN94/97 &bloated stars interaction & sphere  & NGC 4395 does apply \\
        &P88     & SMBH potential & irradiated stellar chromospheres &  \\
        &T89     & SMBH potential & bloated stars & \\
        &K89     & SMBH potential & bloated stars & \\
        &TP02    & stellar atmosphere & ... \\ \hline
Model F &P01     & supernova-QSO wind interactions &   &  \\
        &PD85    &supersonic gas flow interact with obstacles & ... & obstacles: SNexp shell \& strong stellar winds\\
        &this paper& clouds bound by SMBHs & BLR gas from the SF disk & episodic appearance of BLR in 4 phases \\
\hline

\end{tabular}
\tablecomments{All the models involve photoionization as main radiation mechanism.
AD: accretion disk; HIL: high ionization lines; LIL: low ionization lines; PI: photoionization; RDR:
radiation-dominated region; HIM: hot intercloud medium.\\
~~~Referenced.$-$
AN94/97: Alexander \& Netzer (1994, 1997); ALW93: Artymowicz, Lin \& Wampler (1993); B81: Beltrametti (1981);
BM75: Blumenthal \& Mathews (1975); BMS83: Begelman, McKee \& Shields (1983); C80: Capriotti et al. (1980);
C88: Collin-Souffrin et al. (1988); CH01: Collin \& Hure (2001); CH11: Czerny \& Hryniewicz (2011); CR96:
Cassidy \& Raine (1996); E80: Edwards 1980; EBS92: Emmering, Blandford \& Shlosman (1992); K89: Kazanas
(1989); H75: Hills (1975); KMT81: Krolik, McKee \& Tarter
(1981); MC97: Murray \& Chiang (1997); N80: Netzer (1980); N03: Nicastro et al. (2003); P88: Penston (1988);
PD85: Perry \& Dyson (1985); P01: Pittard et al. (2001); RNF89: Rees et al. (1989);
R92: Roos (1992); S77: Shields (1977); S85: Shlosman et al. (1985); SN88: Scoville \& Norman (1988); SR85:
Smith \& Raine (1985); T92: Terlevich et al. (1992); T89: Tout et al. (1989); TP02: Torricelli-Ciamponi \& Pietrini 2002;
 VC02: Vilkoviskij \& Czerny
(2002); W59: Woltjer (1959); WMY85: Wandel, Milgrom \& Yahil (1985); W96: Woods et al. 1996; ZSC94: Zurek et al. (1994).
}
\label{blr_model}
\end{sidewaystable*}

\section{Discussion}

\subsection{Brief comparison with other models}
It is generally thought that BLRs are composed of a two-phase medium (Krolik et al. 1981). Properties
of photoionized cold clouds formed through thermal instability can be predicted and generally agree
with observations. But key issues have long remained in doubt. These include the origin of the hot-phase
gas, the evolution of the hot gas and also of the cold BLR clouds, and the geometry of the BLR. The present
model makes clear predictions about the properties of the BLR and clouds, which are testable from
observations. We now make a brief comparison of the present model with previously existing ones.

Table \ref{blr_model} lists existing models of the BLRs\footnote{There has been a considerable amount of
work on theoretically understanding BLRs. We list here the models that we know of in the published
literature, but there may well be others.}. We divided them into five types in light of their main features.
They are: 1) model type  A employs the outer part of the accretion disk, which is irradiated by the central
part; 2) model type B deals with outflows which are approximately parallel to the surface of the accretion
disk; 3) model type C invokes clouds bound by magnetic fields which are poloidal perpendicular to the disk;
4) model type D employs gas stirred by stars in a central cluster; 5) model type E uses bloated stars; and
6) model type F deals with supernova explosions. We briefly compare them with the present model.

First of all, {\em all} models of types A-E are neither able to specify how the BLR gas originates, nor
connect the metallicity in the BLR with feeding the SMBH. Model A employs the outer part of the accretion
disk (e.g. Collin-Souffrin 1987), which is irradiated by the central engine. Clouds may be formed in the
outer region of the accretion disk, but the origin of the high metallicity in the BLR remains open. These
models could explain the dependence of the BLR on the accretion rates, but the detailed relationship with
the accretion disk is highly uncertain. On the other hand, the presence of non-BLR AGNs with normal Eddington
ratios (i.e. the Panda AGNs) strongly challenges Model A and also the improved suggestions of Gu \& Huang
(2002), Nicastro et al. (2003) and Elitzur \& Ho (2009). The same is true with models of type B and C.
Interaction between stars and the accretion disk has been employed by model D in Artymowicz et al. (1993)
and Zurek et al. (1994) and leads to production of metals. However, these models do not attempt to explain
the establishments of the BLR geometry or the dynamics of clouds. Model
E invokes bloated stars, but it is hard to explain a clear trend relating metallicity and luminosity of AGNs.
In particular, there is no space for the huge size of the bloated stars (Laor et al. 2006).

Second, properties of clouds in the models A-F are only deduced from the necessary conditions for
photoionzation in light of the cooling and heating function or from observed features. The present model
is capable of predicting cloud properties, dynamics of clouds and how they evolve. This is the advantage.
In addition, these are stationary models of the BLR clouds, with very short lifetimes against destruction
through dynamical friction with their surroundings. The existing models at best assume some motion of clouds
around the SMBH in order to predict the line profiles (e.g. Netzer \& Marziani 2010). In contrast to these
previous models, the present model predicts the details of the BLR's evolution with time, giving rise to
the episodic appearance and reappearance of BLRs during the AGN lifetime. The dynamics of clouds
are fully predicted, and are different in phases II and III. The dynamical destruction of clouds is
explicitly followed and plays a key role in driving the evolution.

Third, in our model the BLR geometry and dynamics are both fully predicted. They are naturally derived from
the dynamics of the CAS, the birthplace of the clouds, The spatial distribution of BLR clouds follows the CAS
density profile, but also the initial angular momentum of clouds tracks the rotation of the CAS. The CAS
properties in turn follow predictably from the properties of the SF disk. In most other models the geometry
is not known, so cannot be tested. We have not yet carried out the numerical simulation of the reverberation
mapping relation for the present model, but that needs to be done in the future and compared in detail with
observations.

Fourth, the present model suggests the HIL regions could develop outflows through enhanced radiation pressure
and predict a potential connection between the BLR and the NLR. As a consequence of the connection, the
outflowing clouds could evolve into warm absorbers appearing in soft X-ray bands. We expect an intrinsic
relation between high ionization lines and the warmer absorbers. Outflowing clouds are also employed in other
models to explain the BLRs. Spherical outflows (Beltrametti 1981) and
ascending winds from the accretion disks (Shlosman et al. 1985) have been investigated. Emmering et al.
(1992) studied magnetized outflows perpendicular to the disk whereas Murray \& Chiang (1995) and
Chiang \& Murray (1996) considered
flows parallel to the disk. Though these outflow models can generally generate the profiles of broad emission
lines, the geometry is neither certain, nor do they incorporate the stratified structure of the BLRs.

The advantages of the present model can be summarized as follows:
\begin{itemize}
\item The model suggests that BLR formation is a by-product of feeding SMBHs. Without supernova
explosions driving outward transportation of angular momentum,
neither accretion onto the SMBH nor BLR formation are able to proceed. This leads to a natural
relationship between high metallicity and accretion onto the SMBH.

\item Cold clouds are produced in an evolving Compton gas originated from the star forming disk,
avoiding the crisis of the dynamical destruction of clouds, which has not been solved in the
classical model. On the contrary, the episodic appearance of the BLR employs the dynamical destruction of
clouds, which is a necessary element in the present model.

\item The geometry and dynamics of clouds follows the Compton gas. The BLR is naturally divided into the
low and high ionization regions related with different phenomena of emission lines. Outflows from the
HIL regions bridge the BLR and narrow line regions separated by 5 orders in size scale.

\item The model predicts a new kind of AGNs, which have higher Eddington ratio, but do {\em not}
have BLRs. This enriches the classes of known AGNs.

\item The model predicts a metallicity gradient between high and low ionization line regions.
This is one of the main characteristics of the present model.

\end{itemize}

The present model has further implications to observations beyond the scope of BLRs as a by-product
of feeding the SMBHs. The fact that the BLRs emit more energy than they received from UV and X-rays,
which is known as ``energy budget problem" (Collin-Souffrin 1986; Netzer 1990) could be explained
by supplementing the UV emission with light from the massive stars in the star forming disk provided
the star formation rates are high enough (see \S6.2). If this is true, quasars with the energy budget
problem (Netzer 1985) should have higher star formation rates, and hence higher metallicity. This possibility should
be explored.

In this paper, we show that broad line regions are formed as a consequence of the accretion flow onto
the SMBHs. Since the evolving BLRs are not stationary during the AGN lifetime, the BLR will be renewed
in the next episode with a duty cycle of a few $10^3$yrs. The SMBH activities are composed of: 1) BLR
episodes and 2) AGN episodes. This indicates that the BLRs will have similar properties in low- and
high-high redshift quasars. There is no measurable cosmological evolution of spectral energy distribution
(Brandt et al. 2007) or of metallicity (Hamnann \& Ferland 1993; Warner et al. 2004; Shemmer et al. 2004;
Matsuoka et al. 2011). Also as relics of SMBH activities, stellar rings/belts and eventually nuclear
compact star clusters (Wang et al. 2010) should be produced during random accretion onto SMBHs (King \&
Pringle 2007) as evidenced by the results from $\eta-$equation of SMBH spins (Wang et al. 2009). A future
paper will provide a global explanation of fueling SMBH, BLRs, metallicity,
nuclear compact star cluster and evolution of SMBH as they spin down.

\subsection{Future work}
First of all, we approximate the Compton gas as barotropic gas in a quasi-steady state for discussion of
the thermal instability of the Compton gas. This approximation holds provided the supplying timescale is
longer than the thermal instability timescale. The present model should be improved for some special cases
with fast injection from the SF disk. We simply discuss the fate of the Compton gas with several critical
densities. We neglect the detailed contraction of clouds. Second, we do not discuss the detailed dynamics
of clouds after their formation from the Compton gas. For simplicity, we simplify their motion to be a
spiral path down to the SF disk with the initial angular momentum of their birth. This is important for
profiles from the bound clouds in BLRs. Third, the dynamics of outflowing clouds is not discussed with
regard to the connection between the BLR and NLR. We plan to investigate the evolution
of the outflowing clouds in a self-consistent way, including expansion, involving the swept medium and
metallicity mixture, for the connection between the BLR and the NLR. These processes should produce
multiwavelength features. Warm absorbers in X-rays should somehow be linked with the clouds in BLR-I.
This could be examined by comparing the properties of UV \civ\, profiles with the soft X-ray spectrum.

One may ask about two other potential energy sources for heating the CAS. The first is self-heating when
the CAS is accreted, through viscosity resembling the $\alpha-$prescription. This is similar
to the case studied by Meyer \& Meyer-Hofmeister (1994).
The energy flux due to the viscosity is given by $F_{\rm CAS}=3G\mbh\dot{M}_{\rm CAS}/4\pi R^3$,
where $\dot{M}_{\rm CAS}$ is the accretion rates of the CAS onto the SMBH. The energy flux from
the central engine reads $F_{\rm AD}=\eta \dotmbh c^2/4\pi R^2$. We have
$F_{\rm CAS}/F_{\rm AD}=1.5\times 10^{-3}r_3^{-1}(\dot{M}_{\rm CAS}/\dotmbh)\ll 1$. However, the
leakage of the CAS gas through accretion would play an important role in the evolution of the CAS. Actually
the accretion rate in the CAS is much less than the injection rate from the SF disk. This guarantees
the results presented in this paper. The second possible additional energy source is radiation from the
star-forming disk. The energy flux from the SF disk is
$F_{\rm SF}=\eta_*\dot{\Sigma}_*c^2$, where $\eta\sim 10^{-3}$ for a Salpeter initial mass function
(Thompson et al. 2005) and $\dot{\Sigma}_*$ is the surface density of the star formation rate. We
have $F_{\rm SF}/F_{\rm AD}=(\eta_*\eta^{-1})(\dot{{\cal R}}_*/\dotmbh)\sim 0.1$ if
$\dot{{\cal R}}_*\sim 10 \dotmbh$ and $\eta=0.1$. So this energy supply to the CAS could be important
in some cases. Especially, its contribution may change the Compton temperature, but this depends on its
SED (e.g. Sirko \& Goodman 2003). Future work should include the leakage of the CAS mass through
accretion onto the SMBH and the heating of the SF disk itself.

Accurate cooling functions should be used to determine the new thermal equilibrium to obtain
the final column density and then the dynamics of clouds in the radiation fields. Furthermore,
we may, in principle, include other heating mechanism discussed in Krolik et al. (1981).
The dynamics of the clouds should be specified more completely by including the friction due to the CAS,
radiation pressure, and even evaporation of clouds and condensations.
This will allow us to study the detailed processes of the evolution of clouds.

Composite spectra from the assembly clouds in the BLRs should be created by adding all the spectra from
every individual clouds (Ferland \& Baldwin 1999). Line profiles from such a complicated model will be calculated
in light of the cloud's dynamics.
Clouds in LIL regions only have one orbital lifetime in phase II, and an even shorter lifetime
in phase III. This complicates the calculations of the line profiles, as compared to the scheme used by
Netzer \& Marziani (2010). Reverberation mapping of broad emission line regions given in this
paper should be done in future. We will carry out these results in a separate paper.

The present model will also need modifications to be able to properly describe those low luminosity AGNs
which lack BLRs. For a very low accretion rate, the SF disk will be so thin that the SNexp will break out
and develop strong expansion outside of the disk. The central engines of these low-luminosity objects are
expected to develop advection-dominated accretion flows (ADAF) with winds that will collide with the
supernovae winds, producing significant $\gamma-$ray emission. Such objects should manifest themselves as
low ionization emission line regions (LINERs). The predicted $\gamma-$rays may be observable with {\em Fermi}
or {\em HESS}. A detailed model for this scenario will be developed separately.

\section{Conclusions}
We have developed a model that describes the way in which a star-forming region in the outer parts of an
AGN accretion disk (beyond the self-gravitation radius) will quite naturally, just as an inevitable result
of the accretion process, lead to a two-component broad emission line region. We have shown that the BLR
should appear, disappear, and then reappear in a continuing series of episodes. Within each episode, the
BLR will progress through an evolutionary cycle consisting of four phases. This will lead to an observable
sequence of quasar broad emission line spectral types, with predictable characteristics. We have briefly
tested these predictions against observations. This episodic model for the BLR predicts: 1) the observed
result that  $16\%$ of SDSS quasars show systematic redshifts correlated with the Eddington ratios whereas
about $80\%$ quasars show overlapping profiles of intermediate and broad components; 2) a new class of bright
type II AGNs without broad emission lines and with high Eddington ratios; 3) an intrinsic connection of broad
and narrow line regions through the outflows developed from the high ionization regions; and 4)
a significant difference of metallicity between the high and low ionization regions. We have described
mostly-published observational evidence suggesting that each of these predictions is correct.

The broad line regions are naturally divided into high- and low-ionization (HIL and LIL) regions. The HIL
region is characterized by outflows. The LIL region has a mass circulation with the SF disk and, later, a
mass cycle with the Compton atmosphere in phase II and III, respectively. Properties of clouds are
self-consistently obtained from the model. An important advantage of the present model is that it employs
clouds which are not in stationary states, avoiding the difficulties encountered by models in which dynamical
destruction of the clouds must be prevented.

The striking predictions of our model have been compared to SDSS data in a preliminary way. Further
observational tests should in the future provide a more detailed test of this transient BLR model.

\acknowledgements{We appreciate the stimulating discussions among the members of IHEP AGN group.
JMW is grateful to H.-Y. Zhou for interesting conversations about the non-BLR quasars. Y.-R. Li
is thanked for plotting the cartoon of broad line regions. The research is supported by NSFC-10733010
and -10821061, and 973 project (2009CB824800). JAB and GJF are grateful to J.-M. Wang and the Institute
of High Energy Physics of the Chinese Academy of Sciences for their hospitality. JAB acknowledges support
by NASA grant NNX10AD05G,  and GJF by NSF (0908877; 1108928 and 1109061), NASA (07-ATFP07-0124;
10-ATP10-0053; and 10-ADAP10-0073), JPL (RSA No. 1430426) and STSci (HST-AR-12125.01 and HST-GO-12309).
}

\appendix

\section{An isothermal Compton atmosphere}
As we shown in Figure 3, the CAS is not an isothermal atmosphere. However, it could be a good approximation,
providing very simple results. In this appendix, we discuss the isothermal CAS. For $\gamma=1$, Equations
(40) and (41) yield solution as
\begin{equation}
K_0\ln\rho_0+\Phi=\int \Omega^2 RdR,
\end{equation}
where $K_0=kT_{\rm Comp}/m_p$ and equation (46) can be recast as
\begin{equation}
\rho_0(r,h)=\displaystyle \exp\left\{\frac{m_pc^2}{kT_{\rm Comp}}\left[\frac{\omega_0^2r_{\rm out}^{2q-3}}{4(1-q)}\frac{1}{r^{2(q-1)}}
            +\frac{1}{2(r^2+h^2)^{1/2}}+C_0\right]\right\}.
\end{equation}
The results are numerically similar to that as shown in Figure 6 and 7 and we thus omit to carry them
out. It should be noted that the index $\gamma\neq 1$ in the development of thermal instability for
an isothermal CAS.

\section{Dispersion relation}
The thermal instability of a rotating corona above a galactic disk has been studied by Binney, Nipoti \&
Fraternali (2009) and Nipoti (2010), however the results of his analysis cannot directly be scaled down to
the BLR case. Following Nipoti (2010), the perturbation equations (48-52) give
\begin{equation}
\left[\begin{array}{lllll}
i\kr\rho_0        & i\kz\rho_0     &  0            & -i\homega     & 0     \\
-i\homega \rho_0    &   0            & -2\Omega\rho_0& -\apr c_0^2 & i\kr  \\
0                 &-i\homega\rho_0 & 0             &  -\apz c_0^2& i\kz  \\
\Omega+\Omega_R   & \Omega_z       & -i\homega     &  0          & 0     \\
\apr-\gamma\arhor & \apz-\gamma\arhoz &0 &(i\homega-\omega_d)\gamma\rho_0^{-1}& -i\homega p_0^{-1}\end{array}\right]
\left[\begin{array}{c}
\vr^1\\ \vz^1\\ \vphi^1\\ \rho_1\\ p_1\end{array}\right]=
\left[\begin{array}{c}
0 \\ 0\\ 0\\ 0\\ 0\end{array}\right],
\end{equation}
where $\omega_{\rm d}=\omega_{\rm c}+\omega_{\rm th}$, $\Omega_z=\pp (R\Omega)/\pp z$ and
$\Omega_R=\pp (R\Omega)/\pp R$. Here $\apr=\left(\pp p_0/\pp R\right)/p_0$ and
$\apz=\left(\pp p_0/\pp z\right)/p_0$ are the inverse of the pressure scalelength and scaleheight,
respectively; $\arhor=\left(\pp \rho_0/\pp R\right)/\rho_0$ and $\arhoz=\left(\pp \rho_0/\pp z\right)/\rho_0$
are the inverse of the density scalelength and scaleheight, respectively.
The non-zero solution of the above equations gives
\begin{equation}
\left|\begin{array}{lllll}
i\kr\rho_0        & i\kz\rho_0     &  0            & -i\homega     & 0     \\
-i\homega \rho_0    &   0            & -2\Omega\rho_0& -\apr c_0^2 & i\kr  \\
0                 &-i\homega\rho_0 & 0             &  -\apz c_0^2& i\kz  \\
\Omega+\Omega_R   & \Omega_z       & -i\homega     &  0          & 0     \\
\apr-\gamma\arhor & \apz-\gamma\arhoz &0 &(i\homega-\omega_d)\gamma\rho_0^{-1}& -i\homega p_0^{-1}\end{array}\right|=0,
\end{equation}
yielding
\begin{equation}
a_5\hatn^5+a_3\hatn^3+a_2\hatn^2+a_1\hatn+a_0=0,
\end{equation}
where $\hatn=-i\hat{\omega}$, the coefficients are
\begin{equation}
a_5=-\frac{\rho_0}{p_0},
\end{equation}
\begin{equation}
a_3=-\gamma(\kr^2+\kz^2)-i\gamma(\kr\arhor+\kz\arhoz)-\frac{2\rho_0\Omega}{p_0}(\Omega+\Omega_R),
\end{equation}
\begin{equation}
a_2=-\gamma(\kr^2+\kz^2)\omega_{\rm d}=-\gamma k^2\omega_{\rm d},
\end{equation}
\begin{equation}
a_1=a_{11}+a_{12}+a_{13}+(a_{14}+a_{15})i,
\end{equation}
and
\begin{equation}
a_0=2\gamma\kz\Omega\omega_d\left[\kr\Omega_z-\kz(\Omega+\Omega_R)\right].
\end{equation}
where
\begin{equation}
\begin{array}{l}
a_{11}=c_0^2\left(\kr\apz-\kz\apr\right)^2\sim c_0^2(kR^{-1})^2,\\ \\
a_{12}=c_0^2\gamma\left(\kz\arhor-\kr\arhoz\right)\left(\kr\apz-\kz\apr\right)\sim c_0^2\gamma (kR^{-1})^2,\\ \\
a_{13}=2\gamma\kz\Omega\left[\kr\Omega_z-\kz(\Omega+\Omega_R)\right]\sim 2\gamma k^2\Omega^2,\\ \\
a_{14}=-2\gamma\Omega\kz\left[\arhoz(\Omega+\Omega_R)-\arhor\Omega_z\right]\sim 2\gamma \Omega^2(kR^{-1}),\\ \\
a_{15}=2\Omega\Omega_z(\kr\apz-\kz\apr)\sim 2\Omega^2(kR^{-1}).
\end{array}
\end{equation}

For low-frequency perturbations, we can use the approximations
\begin{equation}
\omega^2\ll c_0^2(\kr^2,\kz^2); ~~~~{\rm and}~~
(\Omega^2,\Omega_z^2,\Omega_R^2)\ll c_0^2(\kz^2,\kr^2),
\end{equation}
whereas for the short-wavelength case we have the approximations
\begin{equation}
(|\kr|, |\kz|)\gg(|\arhor|,|\apz|,|\apz|,|\apr|).
\end{equation}
The fifth order term vanishes in the dispersion relation.

The terms of $a_3$ are of order $|a_{31}|=\gamma(\kr^2+\kz^2)\sim \gamma k^2$,
$|a_{32}|=\gamma(\kr\arhor+\kz\arhoz)\sim \gamma kR^{-1}$ and
$|a_{33}|=2\rho_0\Omega(\Omega+\Omega_R)/p_0\sim \Omega^2/c_0^2$. We find that
$|a_{31}/a_{32}|\sim kR\gg 1$ and $|a_{31}/a_{33}|\sim \gamma k^2c_0^2/2\Omega^2\gg 1$.
We then have the simplified form
\begin{equation}
a_3=-\gamma(\kr^2+\kz^2)=-\gamma k^2.
\end{equation}
The coefficients of $a_{1i}$ are of $a_{11}\sim a_{12}\sim a_{13}$, and $a_{13}\gg (a_{14},a_{15})$, we have
\begin{equation}
a_1=a_{11}+a_{12}+a_{13}.
\end{equation}

The coefficients can be simplified through the operator introduced by Balbus (1995), which is defined by
\begin{equation}
\cald=\frac{\kr}{\kz}\frac{\pp}{\pz}-\frac{\pp}{\pr}.
\end{equation}
We use the following relations for convenience
\begin{equation}
\kr\apz-\kz\apr=\kz\cald \ln p_0; ~~~~~\kr\arhoz-\kz\arhor=\kz\cald \ln \rho_0,
\end{equation}
and
\begin{equation}
\cald s_0=\cald \ln p_0-\gamma\cald \ln \rho_0
\end{equation}
where the entropy $s_0=\ln p_0\rho_0^{-\gamma}$. We have other relations
\begin{equation}
\kr\Omega_z-\kz\Omega_R=\kz\cald(R\Omega),
\end{equation}
and
\begin{equation}
\cald(R\Omega)-\Omega=\frac{\cald\left(R^4\Omega^2\right)}{2R^3\Omega}.
\end{equation}
With the help of (B14-B17), we have
$$a_{11}=c_0^2\kz^2\left(\cald \ln p_0\right)^2,$$
$$a_{12}=-c_0^2\gamma\kz^2\left(\cald \ln p_0\right)\left(\cald \ln \rho_0\right),$$
$$a_{13}=\frac{\gamma \kz^2}{R^3}\cald (R^4\Omega^2),$$
$$a_0=\omegad \frac{\gamma \kz^2}{R^3}\cald (R^4\Omega^2),$$
and
\begin{equation}
a_1=c_0^2\kz^2\left(\cald \ln p_0\right)\left(\cald s_0\right)+\frac{\gamma \kz^2}{R^3}\cald (R^4\Omega^2).
\end{equation}

We have the dispersion relation as
\begin{equation}
-\gamma k^2\hatn^3-\gamma k^2\omegad \hatn^2+
\left[c_0^2\kz^2\left(\cald \ln p_0\right)\left(\cald s_0\right)+\frac{\gamma \kz^2}{R^3}\cald (R^4\Omega^2)\right]\hatn
+\omegad \frac{\gamma \kz^2}{R^3}\cald (R^4\Omega^2)=0,
\end{equation}
reducing to
\begin{equation}
\hatn^3+\omegad \hatn^2
-\left[c_0^2\frac{\kz^2}{\gamma k^2}\left(\cald \ln p_0\right)\left(\cald s_0\right)
+\frac{\kz^2}{k^2R^3}\cald (R^4\Omega^2)\right]\hatn
-\omegad \frac{\kz^2}{k^2R^3}\cald (R^4\Omega^2)=0.
\end{equation}
Introducing the Brunt-V\"ais\"al\"a frequency
\begin{equation}
\omega_{\rm BV}^2=-c_0^2\frac{\kz^2}{\gamma k^2}\left(\cald \ln p_0\right)\left(\cald s_0\right)
                 =-\frac{\kz^2}{k^2}\frac{\cald p_0}{\gamma \rho_0}\cald s_0,
\end{equation}
and the differential rotation term
\begin{equation}
\omega_{\rm rot}^2=-\frac{\kz^2}{k^2R^3}\cald (R^4\Omega^2)=-\frac{\kz^2}{k^2R^3}\cald \ell^2,
\end{equation}
where $\ell=\Omega R^2$ is the specific angular momentum, we have the dispersion equation as
\begin{equation}
\hatn^3+\omegad \hatn^2+\left(\omega_{\rm BV}^2+\omega_{\rm rot}^2\right)\hatn+\omega_{\rm rot}^2\omega_{\rm d}=0.
\end{equation}
The simplified version of equation (B22) for barotropic gas reads
\begin{equation}
\omega_{\rm BV}^2=\frac{\kz^2}{k^2}\frac{c_0^2\cala_{pz}^2}{\gamma}\left(\frac{\gamma}{\gamma^{\prime}}-1\right)
                  \left(\frac{\kr}{\kz}-\frac{\cala_{pR}}{\cala_{pz}}\right)^2,
\end{equation}
where we use the assumption that the pressure is only function of density, and we define
\begin{equation}
\gamma^{\prime}\equiv \frac{d\ln p_0}{d\ln \rho_0},
\end{equation}
which can be regarded as a local polytropic index. For the overall barotropic gas, $\gamma^{\prime}=\gamma$, we
have $\omega_{\rm BV}^2=0$, and the dispersion equation reads
\begin{equation}
(\hatn^2+\omega_{\rm rot}^2)(\hatn+\omega_{\rm d})=0,
\end{equation}
with the solution
\begin{equation}
\hatn^2=-\omega_{\rm rot}^2~~~~{\rm or}~~~~\hatn=-(\omega_{\rm c}+\omega_{\rm th}).
\end{equation}

\clearpage

\renewcommand{\baselinestretch}{0.65}

\end{document}